\documentclass[a4]{spie}  
\usepackage[]{graphicx, epstopdf}
\usepackage{float}
\usepackage{subfig}
\usepackage{amsmath}
\usepackage{amssymb}
\usepackage{fixltx2e}

\newcommand{\LeftEqNo}{\let\veqno\@@leqno}

\title{Focal plane wavefront sensing and control strategies for high-contrast imaging on the MagAO-X instrument} 

\author{{\bf \small Kelsey Miller\supit{a,b}, Jared R. Males\supit{a}, Olivier Guyon\supit{a,b,c,d}, Laird M. Close\supit{a}, David Doelman\supit{e}, Frans Snik\supit{e}, \\Emiel Por\supit{e}, Michael J. Wilby\supit{e}, Chris Bohlman\supit{a}, Jennifer Lumbres\supit{a,b}, Kyle Van Gorkom\supit{a,b}, Maggie Kautz\supit{a,b} Alexander Rodack\supit{a,b}, Justin Knight\supit{a,b}, Nemanja Jovanovic\supit{f}, Katie Morzinski\supit{a}, Lauren Schatz\supit{a,b}}
\skiplinehalf
\small
\supit{a}University of Arizona, Steward Observatory, Tucson, Arizona, United States \\
\supit{b}University of Arizona, College of Optical Sciences, Tucson, Arizona, United States \\
\supit{c}National Institutes of Natural Sciences, Subaru Telescope, National Observatory of Japan, Hilo, Hawaii, United States \\
\supit{d}National Institutes of Natural Sciences, Astrobiology Center, Mitaka, Japan \\
\supit{e}Leiden Observatory, Leiden University, Leiden, The Netherlands \\
\supit{f}Caltech Optical Observatory, California Institute of Technology, 1200 E. California Blvd, Pasadena, CA 91125, USA \\
}
\normalsize

\begin{document} 
  \maketitle 




\begin{abstract}
The Magellan extreme adaptive optics (MagAO-X) instrument is a new extreme adaptive optics (ExAO) system designed for operation in the visible to near-IR which will deliver high contrast-imaging capabilities.  The main AO system will be driven by a pyramid wavefront sensor (PyWFS); however, to mitigate the impact of quasi-static and non-common path (NCP) aberrations, focal plane wavefront sensing (FPWFS) in the form of low-order wavefront sensing (LOWFS) and spatial linear dark field control (LDFC) will be employed behind a vector apodizing phase plate (vAPP) coronagraph using rejected starlight at an intermediate focal plane.  These techniques will allow for continuous high-contrast imaging performance at the raw contrast level delivered by the vAPP coronagraph ($\sim$ 6 x 10$^{-5}$). We present simulation results for LOWFS and spatial LDFC with a vAPP coronagraph, as well as laboratory results for both algorithms implemented with a vAPP coronagraph at the University of Arizona Extreme Wavefront Control Lab.
\end{abstract}

\section{Introduction}
Operating in the visible to near-IR, MagAO-X will deliver high Strehl ratios ($\geq$ 70\% at H$\alpha$), high angular resolution performance (14 - 30 mas) and high-contrast imaging ($\leq$$10^{-4}$) between $\sim$1 and 10 $\lambda$/D.\cite{Males2017_MagAOX_PDR}  These capabilities will allow for the study of early stages of planet formation, high spectral-resolution images of stellar surfaces, and the potential for taking the first high-contrast direct images of an exoplanet in reflected light.  The main high-order wavefront sensor (WFS) will be  a state-of-the-art PyWFS; however, to mitigate the impact of static, quasi-static, and NCP aberrations, two focal plane wavefront sensing strategies will be employed: LOWFS and spatial LDFC.  The contrast achieved in the dark hole is highly sensitive to minute aberrations in the optical path, and traditional methods of wavefront sensing such as the PyWFS being deployed on MagAO-X remain blind to NCP and quasi-static aberrations.  To maintain the high contrast achieved by the system coronagraph, focal plane wavefront sensing FPWFS is therefore necessary.  \\  

For Phases I and II of MagAO-X, the vector apodizing phase plate (vAPP) coronagraph will be deployed to create a high-contrast region, referred to as the dark field (DF) or dark hole.  Taking into account both the designed contrast the vAPP can ideally deliver and the effects of small-scale aberrations due to Fresnel propagation through all optical surfaces within the instrument, the final contrast across the 2 - 15 $\lambda$/D dark hole for MagAO-X will be approximately 6 x 10$^{-5}$\cite{Jhen2017_fresnel}.  The light used by both WFS techniques is separated from the science channel by a binary mask, placed at an intermediate focal plane, which transmits the dark holes to the science camera and reflects the stellar bright field (BF) back to a dedicated WFS camera.  LOWFS uses this rejected starlight to sense quasi-static speckles due to instrumental effects at small spatial separations.  Spatial LDFC senses quasi-static speckles at higher spatial separations due to residual wavefront errors by using only the brightest stellar bright field speckles within the rejected starlight that respond linearly to changes in the wavefront.\cite{Miller2017_JATIS_LDFC}  Since spatial LDFC can sense and control speckles at higher spatial frequencies than LOWFS, it can maintain the high contrast across the full extent of the dark hole.  Employing LDFC after correction by the PyWFS allows for continuous high-contrast imaging performance at the raw contrast level delivered by the coronagraph.  In these proceedings, we present a technical overview and simulation results for both LOWFS and spatial LDFC as FPWFS techniques with a vAPP coronagraph, as well as laboratory results from the University of Arizona Extreme Wavefront Control Lab.  These results demonstrate the ability to recover the raw contrast of the vAPP after the introduction of a $\frac{1}{f^2}$ phase screen with a 27 nm RMS wavefront due to the combined WFS and science channel NCP errors.  This capability will allow for continuous high-contrast imaging performance at the 6 x 10$^{-5}$ level on MagAO-X.   \\

\section{vAPP coronagraph at the UA Extreme Wavefront Control Lab}
\begin{figure}[H]
\centering
\includegraphics[scale=0.30]{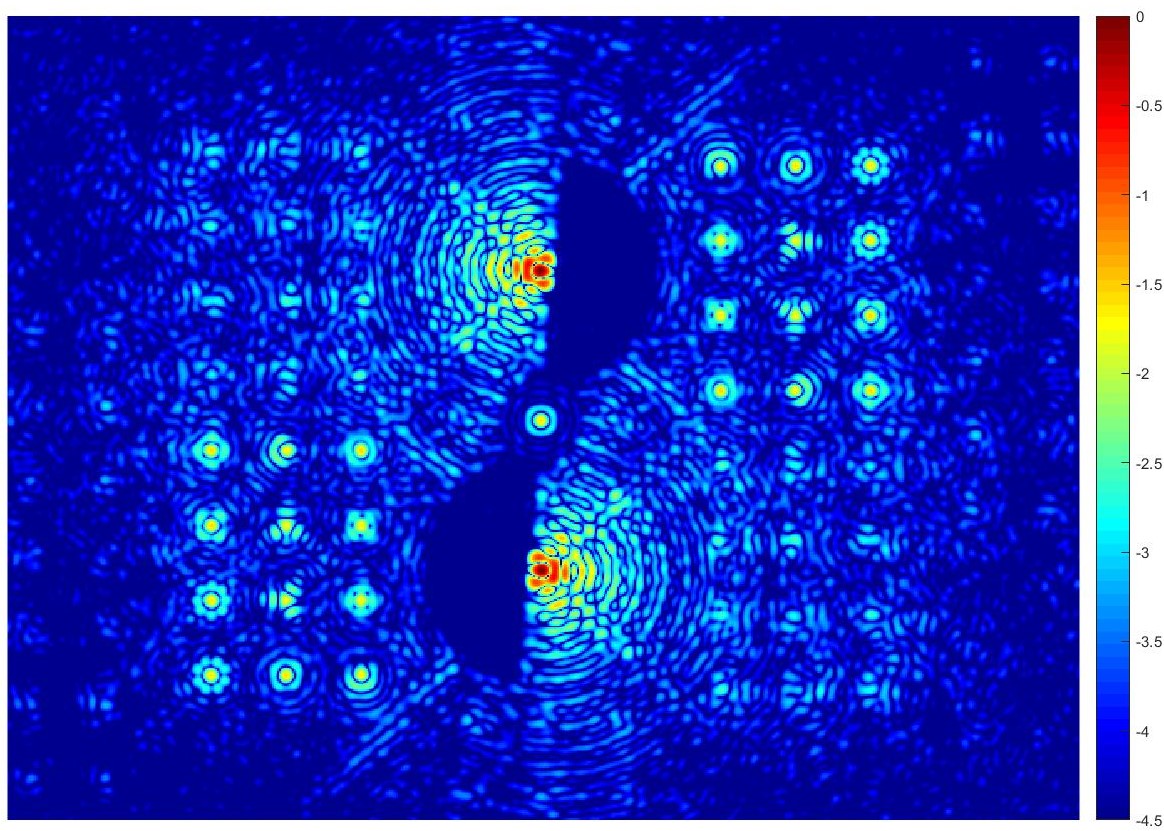}
\caption{Example of the science image delivered by a vAPP coronagraph.  Shown here are two science PSFs each with a 2 - 15 $\lambda$/D dark hole with an average contrast of $\sim$10$^{-5}$.  Further out from the dark holes are twelve MWFS PSFs each encoded with a single low-order Zernike mode.}
\label{fig: zernike_MWFS_sim}
\end{figure}

In the first two phases of the MagAO-X instrument, the system coronagraph will be a vector apodizing phase plate (vAPP) similar to the vAPP coronagraph currently in operation at the Magellan Clay Telescope on the existing MagAO system.\cite{Otten2017_vAPP}  The vAPP is a polarization-dependent coronagraph\cite{Snik2012_vAPP} that creates two polarized copies of the PSF, each with a 180$^{\circ}$ D-shaped dark hole that, when combined, yields a 360$^{\circ}$ high contrast region around the stellar PSF as seen in fig. \ref{fig: zernike_MWFS_sim}.  Between the two polarized science PSFs with dark holes is a copy of the pre-coronagraph system PSF with a peak intensity that is roughly 100x lower than that of the science PSFs.  This PSF is the result of an unpolarized leakage term which is not affected by the vAPP.  Beyond the science PSFs, the vAPP also creates two sets of low-order modal wavefront sensing (MWFS) PSFs, with one set positively biased and one set negatively biased (fig. \ref{fig: zernike_MWFS_sim})\cite{Wilby2017_MWFS}.  These MWFS PSFs are the signal used for driving the LOWFS closed loop, and they can also be used for optical alignment adjustment.  Both uses of the MWFS are addressed in greater detail in Section \ref{sec:LOWFS}. \\

\begin{figure}[H]
\centering
\includegraphics[scale=0.45]{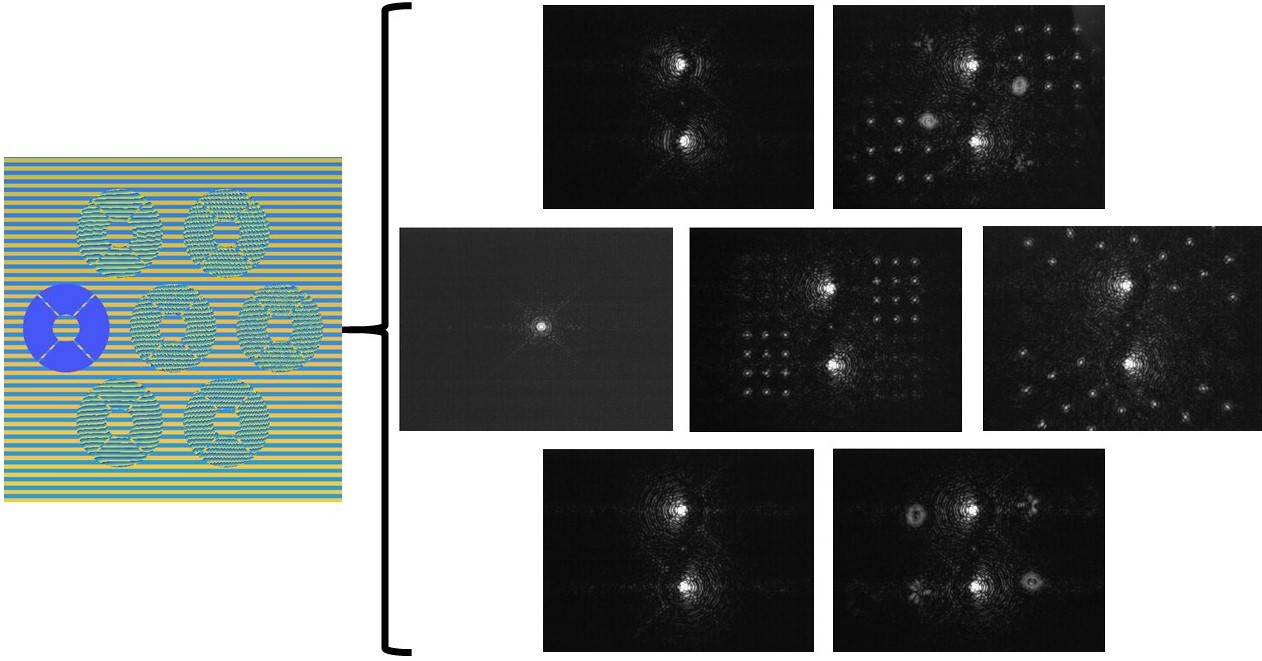}
\caption{The seven vAPP masks and resulting science PSFs, dark holes, and MWFS PSFs on the UA Extreme Wavefront Control Lab's test plate.  $\bf{Top \ row}$:  $\it{(Left)}$  2 - 6 $\lambda$/D dark holes, no MWFS.  $\it{(Right)}$  2 - 11 $\lambda$/D dark holes, 8 Zernikes orthonormalized to the Magellan pupil + phase diversity MWFS.  $\bf{Middle \ row}$:  $\it{(Left)}$  No dark holes, no MWFS.  $\it{(Center)}$  2 - 15 $\lambda$/D dark holes, 12 Zernike MWFS.  $\it{(Right)}$  2 - 11 $\lambda$/D dark holes, 20 Zernikes orthonormalized to the Magellan pupil MWFS.  $\bf{Bottom \ row}$:  $\it{(Left)}$  2 - 11 $\lambda$/D dark holes, no MWFS.  $\it{(Right)}$  2 - 11 $\lambda$/D dark holes, phase diversity MWFS. }
\label{fig: lab_vAPP_PSFs}
\end{figure}

To finalize the design of the MagAO-X vAPP, a plate with seven vAPP masks was designed by Leiden University, manufactured by Imagine Optix, and delivered to the University of Arizona in February 2018 for testing.  Each mask differs by the outer working angle (OWA) of the dark hole and in the MWFS design.  Of the seven masks, six are inscribed with phase patterns producing the two PSFs with dark holes and, in four cases, MWFS PSFs with different modal basis sets.  The seventh mask is a simple binary mask of the Magellan pupil with no phase pattern inscribed.  Of the masks with MWFS PSFs, one creates only phase diversity spots; one contains the first twelve Zernike polynomials; one contains twenty Zernike modes that have been orthonormalized to the Magellan pupil; and one contains eight Zernike modes that have been orthonormalized to the Magellan pupil plus phase diversity spots.  All of these masks and their resulting images can be seen in fig. \ref{fig: lab_vAPP_PSFs}. The design and operation of the MWFS PSFs are described in further detail in Section \ref{sec:LOWFS}.  \\   

\begin{figure}[H]
\centering
\qquad
\subfloat[Reflective Magellan pupil mask elongated along the X axis to match the incoming beam footprint at a 30$^{\circ}$ angle]{\includegraphics[scale=0.044]{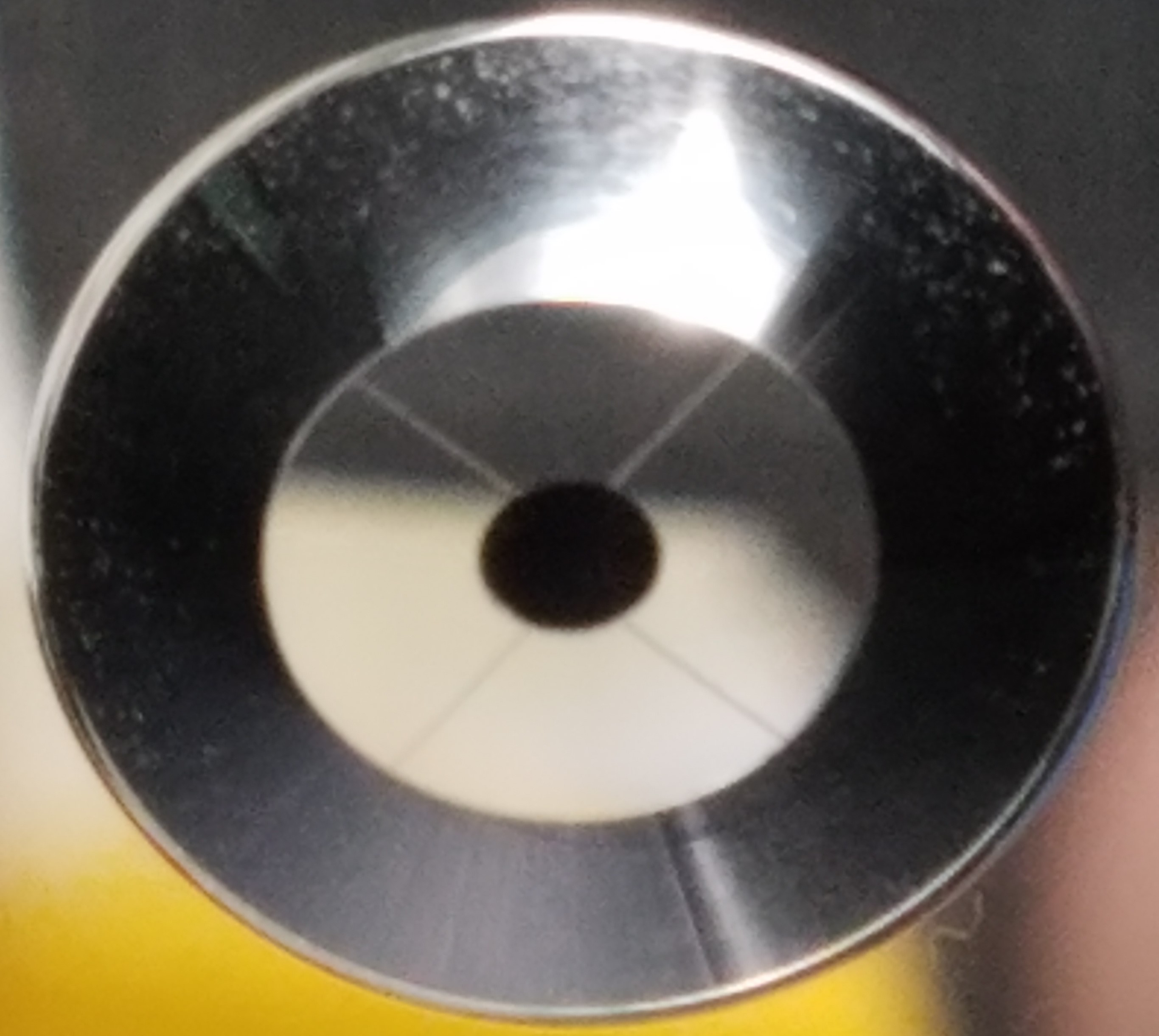}}
\qquad
\subfloat[32 x 32 actuator Boston Micromachines DM]{\includegraphics[scale=0.1]{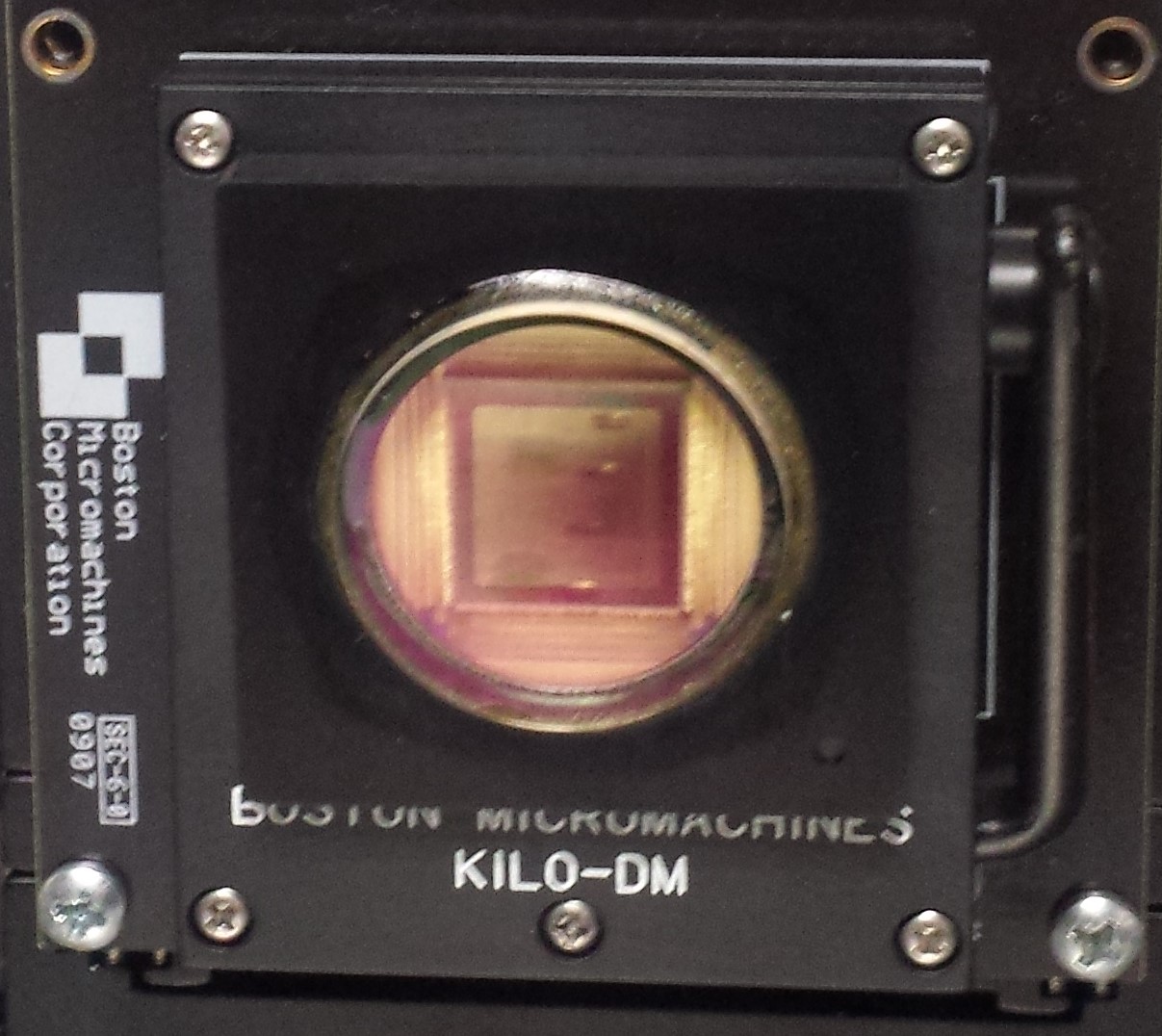}}
\qquad
\subfloat[Mounted transmissive vAPP test plate with the 7 masks manufactured by Imagine Optix]{\includegraphics[scale=0.036]{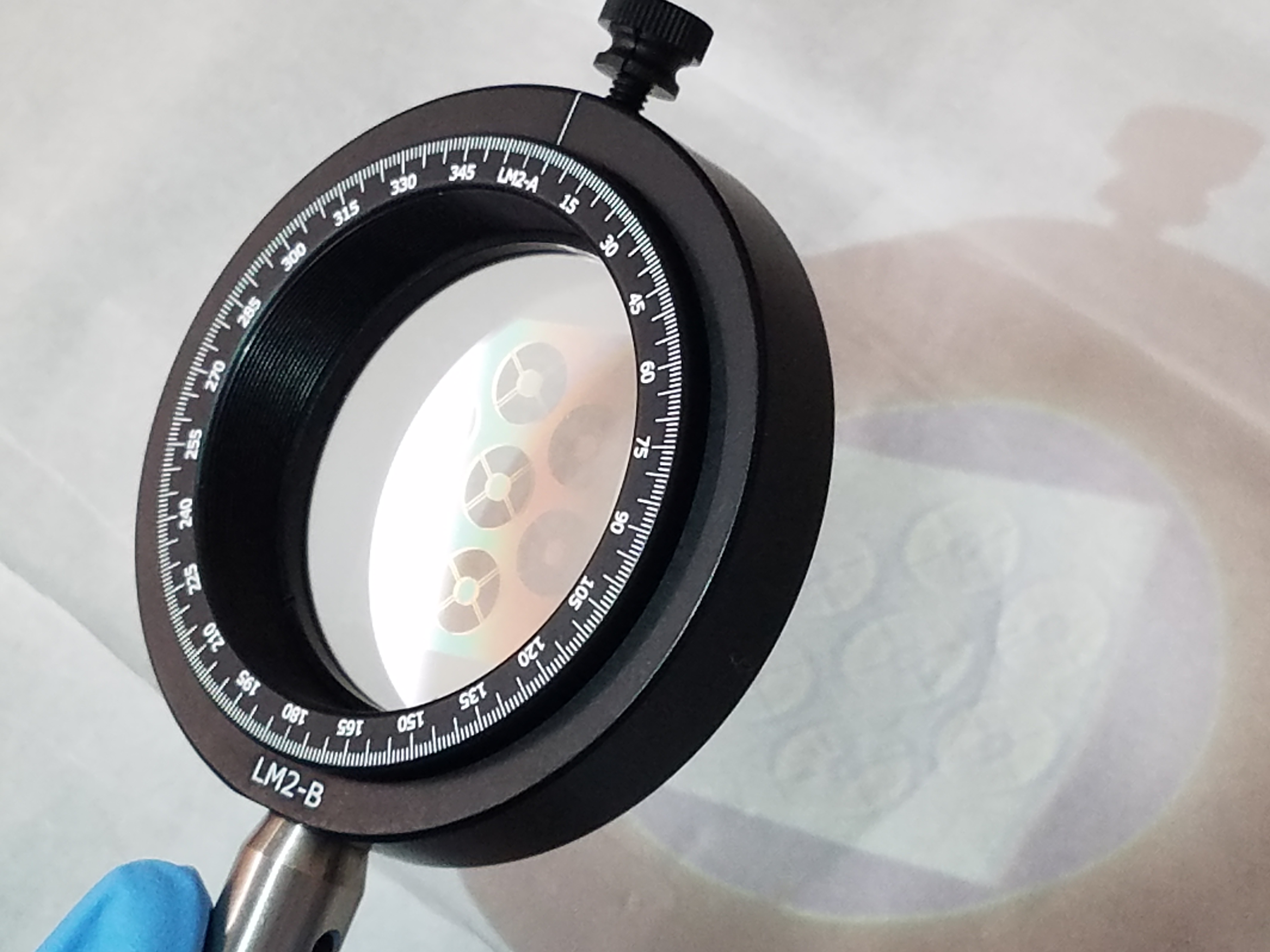}}
\qquad
\subfloat[Full optical layout of the UA Extreme Wavefront Control Lab testbed]{\includegraphics[scale=0.5]{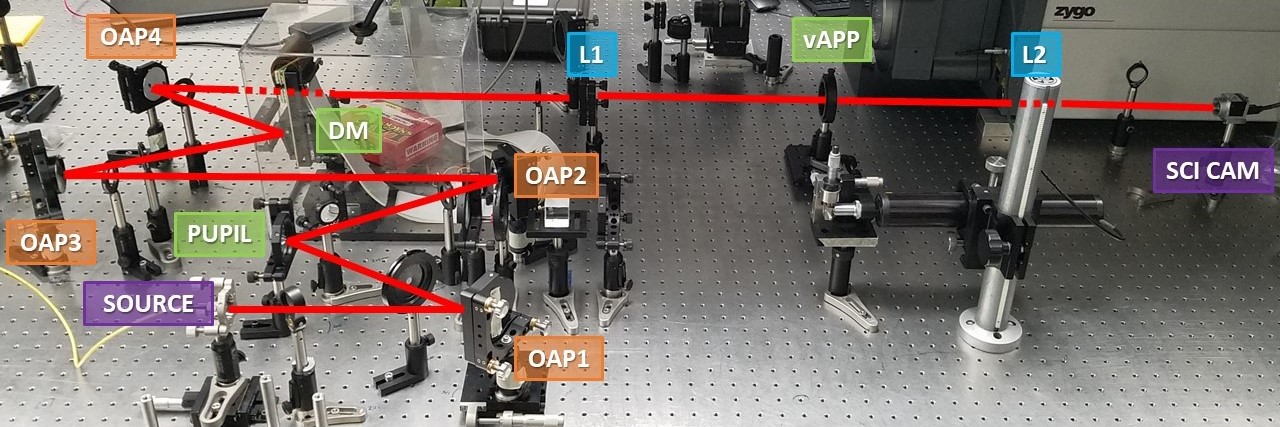}}
\caption{The source injected into the system is a Fianium WhiteLase Micro Supercontinuum laser which is limited by a bandpass filter to pass light only within a 10 nm bandwidth centered at the vAPP design wavelength of 660 nm.  The entrance pupil of the system is defined by a reflective Magellan pupil mask designed with an elongation along the X axis to match the beam footprint seen by the mask when angled at 30$^{\circ}$ with respect to the incoming beam.  This pupil is relayed by off-axis parabolic mirrors (OAPs) to a conjugate pupil plane where the Boston Micromachines Kilo-DM is placed; the DM is used both to inject aberrations into the system and to apply the appropriate correction sensed by the focal plane wavefront sensor.  Another OAP and an achromatic doublet then relay the pupil-DM conjugate plane to the vAPP mask which is mounted on a translation stage to allow for easy mask selection.  This final pupil plane is then brought to focus at the science detector by a single achromatic doublet.}
\label{fig: UA_layout}
\end{figure}

The plate is aligned on the UA Extreme Wavefront Control Lab testbed in a relayed pupil plane conjugate to both a Magellan pupil mask and a Boston Micromachines Kilo-Deformable Mirror (BMC Kilo-DM) (fig. \ref{fig: UA_layout}).  The different masks on the plate were tested both in simulation and on the bench with the intent to determine the regime over which the response of the MWFS PSFs to an applied aberration was linear and the amount of crosstalk between modes.  In both simulation and on the testbed, an aberration was injected into the system using the DM, and the calculated correction was applied using the same DM.  To compare simulated results to testbed results, a model of the testbed was used in simulation which included a scaled model of the BMC Kilo-DM to match the number of actuators across the DM and the elongated shape of the illuminated beam footprint on the testbed DM due to its angled position relative to the incoming beam.  Further explanation of these tests and the results can be found in Sections \ref{sec:LOWFS} and \ref{sec:LDFC}.    \\

\section{Low-order wavefront sensing and control}\label{sec:LOWFS}
Low-order wavefront sensing and control (LOWFS/C) is a well-established technique by which jitter, tip, tilt, and other common low-order aberrations such as coma, astigmatism and defocus are sensed using starlight that has been rejected by the coronagraph.  Traditionally, the signal used to run LOWFS/C in closed loop has been stellar light rejected at either an intermediate focal plane or at the Lyot stop in a conjugate pupil plane in a Lyot coronagraph (known more commonly as LLOWFS/C\cite{Singh2015_LOWFS}) which is then brought to focus at the wavefront sensor camera.  In both of these cases, the light used for wavefront sensing is non-common path with the science beam.  With the vAPP coronagraph, the signal used for closed-loop LOWFS is the MWFS PSFs created in the science image plane, thereby making the wavefront sensor fully common path with the science signal.  With the vAPP, LOWFS/C becomes a focal plane wavefront sensing technique. \\

The MWFS PSFs generated by the vAPP are created by encoding the pupil plane vAPP phase mask with the desired modal basis set.\cite{Wilby2017_MWFS} \cite{Doelman2017_vAPP}.  Each MWFS PSF corresponds to one mode.  Therefore, for example, the twelve Zernike MWFS set seen in fig. \ref{fig: zernike_MWFS_sim} is encoded with the first twelve Zernikes seen below in fig. \ref{fig:zernike_basis}.  \\  

\begin{figure}[H]
\centering
\includegraphics[scale=0.3]{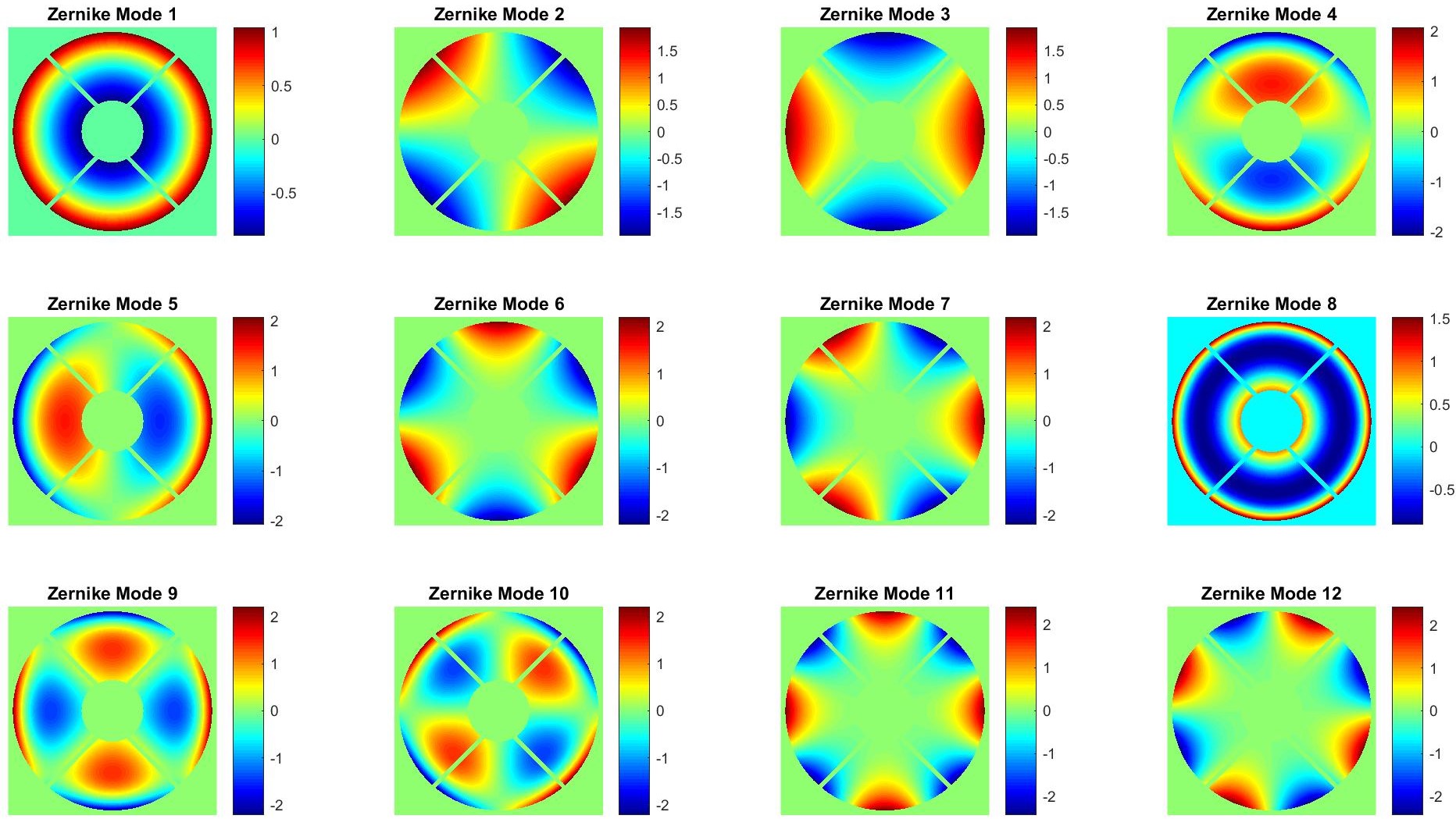}
\caption{The twelve orthonormal Zernike modal basis set masked by the Magellan pupil and encoded on the vAPP mask to create the twelve Zernike MWFS PSFs seen in fig. \ref{fig: zernike_MWFS_sim}.}
\label{fig:zernike_basis}
\end{figure}

\begin{figure}[H]
\centering
\includegraphics[scale=0.4]{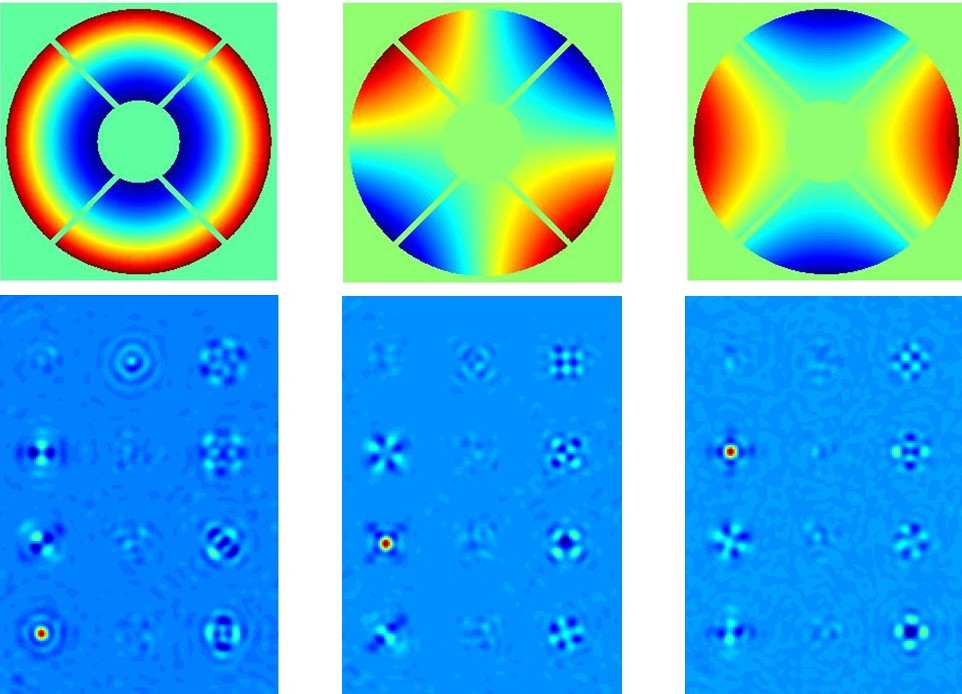}
\caption{The simulated MWFS response to the first three Zernikes used by LOWFS/C, derived by subtracting the negatively biased lower MWFS PSF set of PSFs from the positively biased upper MWFS set. }
\label{fig:zernike_MWFS_example}
\end{figure}

To build the response matrix for LOWFS with the vAPP MWFS, tip and tilt and each mode in the MWFS basis set is applied using the DM, and the response of the MWFS spots is recorded.   The recorded response for each k$^{th}$ mode is the normalized difference in intensity between the positively biased upper MWFS mode (I$_{k+}$) and the negatively biased lower MWFS mode (I$_{k-}$) as dictated by eq. \ref{eq:MWFS_signal}.\cite{Wilby2017_MWFS}  \\
\begin{equation}
I_{k} = \frac{I_{k+} - I_{k-}}{I_{k+} + I_{k-}}
\label{eq:MWFS_signal}
\end{equation}
An example of the first three Zernike modes and the MWFS signal intensity change resulting from their application on the DM in simulation can be seen in fig.\ref{fig:zernike_MWFS_example}.   \\

This method of LOWFS/C using the six different vAPP MWFSs delivered by Leiden University has been tested both in simulation and on the testbed at the UA Extreme Wavefront Control Lab testbed to determine the most efficient MWFS design for the MagAO-X instrument.  Based on preliminary results, a version of the Zernike MWFS design has been selected for the final design.  For this reason, the following work focuses solely on the results of the Zernike MWFS vAPP coronagraph.  \\  
\subsection{LOWFS in simulation}
The first step in testing the Zernike MWFS was the determination of the linear response range of the MWFS.  This was done first in simulation using a model of the vAPP coronagraph and a model of the BMC Kilo-DM in use in the UA Extreme Wavefront Control Lab.  Using the DM, tip, tilt, and each of the twelve Zernike polynomials encoded in the MWFS was applied with an amplitude of 100 nm and the normalized response of the MWFS was recorded as previously described.  This response matrix $\it{G}$ was then inverted and used as the control matrix for the following linearity tests.  \\

\begin{figure}[H]
\centering
\includegraphics[scale=0.45]{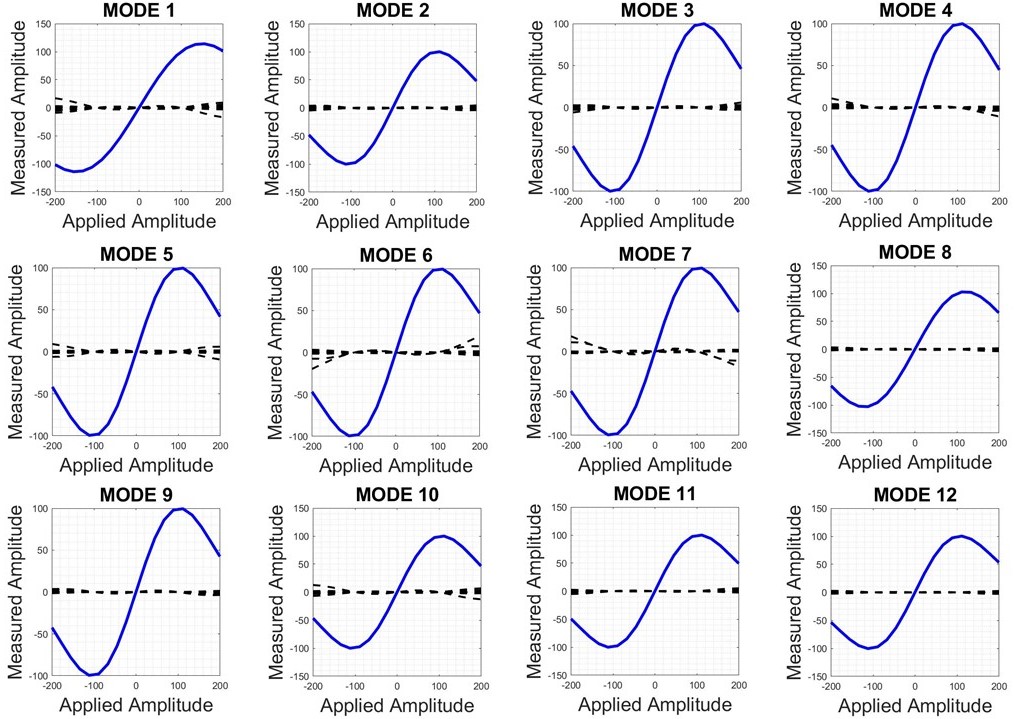}
\caption{Simulated response curves showing the linear response of the MWFS to 12 Zernikes between +/- 100 nm amplitude aberrations.  The blue line represents the response of the mode applied, and the dashed black lines represent the response (or crosstalk) of the 11 other modes to the applied mode.}
\label{fig: linearity_sim}
\end{figure}

To determine the linear response of the MWFS, each aberration was injected into the system with 20 amplitudes ranging from -200 nm to +200 nm with a step size of 22 nm.  The amplitude of all aberrations in the basis set $\it{a}$ was measured for each $\it{k^{th}}$ MWFS mode by fitting the intensity difference image $\it{I}$ to the inverted response matrix $\it{G^{-1}}$ such that 
\begin{equation}
a= G^{-1}I
\label{eq:MWFS_signal}
\end{equation}
The resulting linearity response curves for each aberration are shown in the plots in fig. \ref{fig: linearity_sim}.    \\
As can been seen in the above figure, each mode has a linear, or at least monotonic, response between +/- 100 nm (plotted in blue).  The response of the other modes to the single applied mode or "crosstalk" between the modes is represented by the dashed black lines. It can be clearly seen that, within the linear response regime of each mode, the crosstalk between the other modes is either zero or small enough to be negligible.   It is the combination of this monotonic range and negligible crosstalk between modes that has driven this MWFS's selection for the MagAO-X instrument. \\

\begin{figure}[H]
\centering
\includegraphics[scale=0.6]{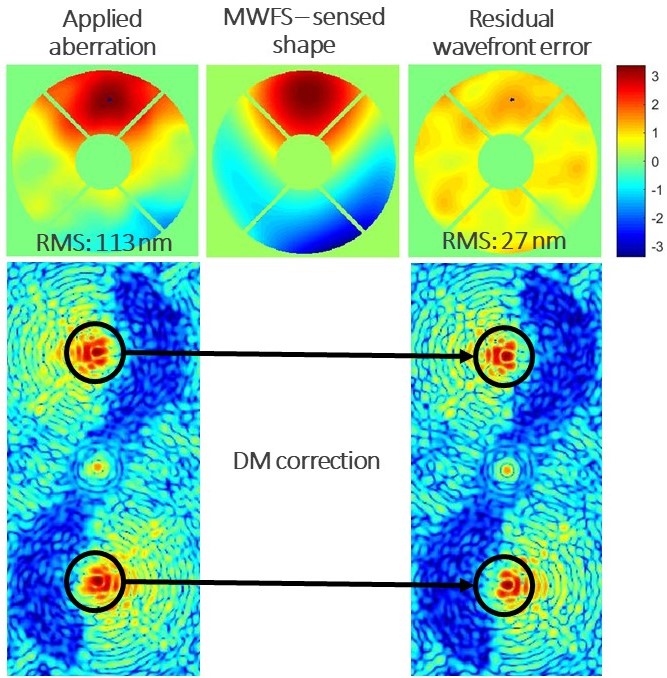}
\caption{LOWFS simulation using 12 Zernike MWFS spots to sense an injected pupil plane aberration and corrected using a model DM.  The simulation converged to a residual RMS of 27 nm from an initial 113 nm in 4 iterations.}
\label{fig: LOWFS_sim_example}
\end{figure}

To ensure its performance in closed-loop, the Zernike MWFS was also tested in simulation.  One example presented here in fig \ref{fig: LOWFS_sim_example}  shows the injection of a pupil plane aberration with an initial RMS of 113 nm.  Using the DM as the corrective element, the simulation converged to a residual wavefront error RMS of 27 nm after 4 iterations.  The same tests were implemented in the lab as well, and the results are presented in the following section.  \\

\subsection{LOWFS laboratory demonstration}

\begin{figure}[H]
\centering
\qquad
\subfloat[Simulated vAPP image]{ \includegraphics[scale=0.25]{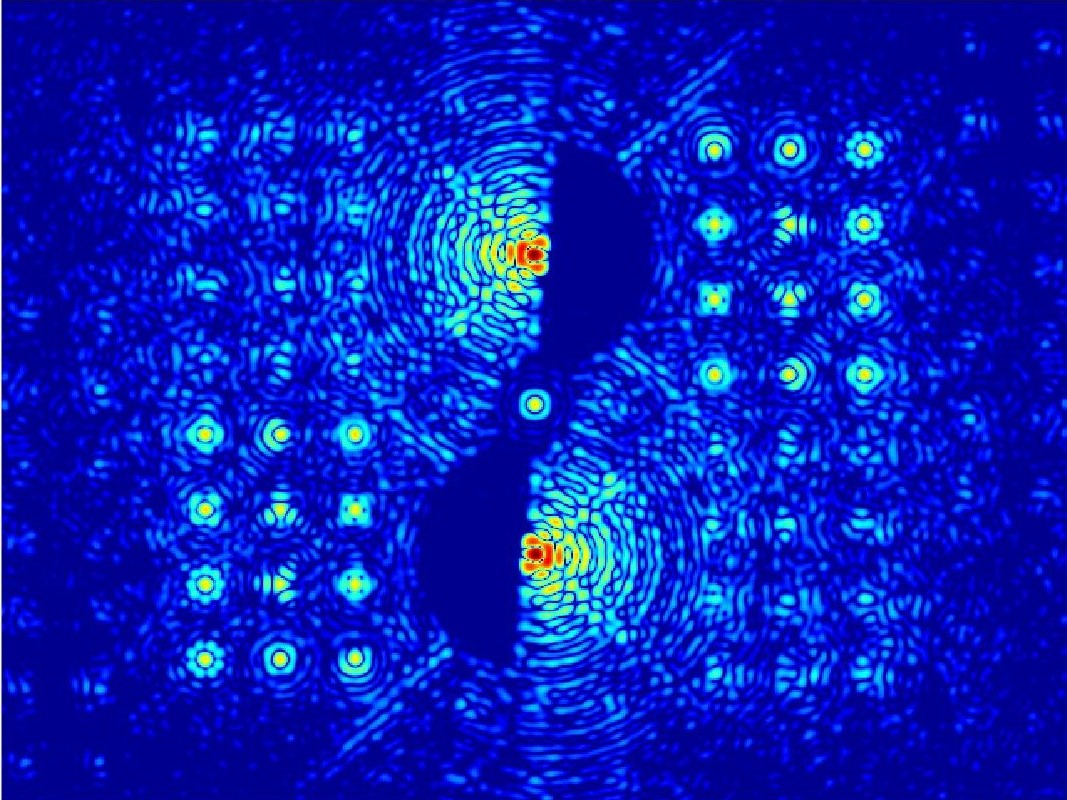}}
\qquad
\subfloat[Laboratory vAPP image]{\includegraphics[scale=0.25]{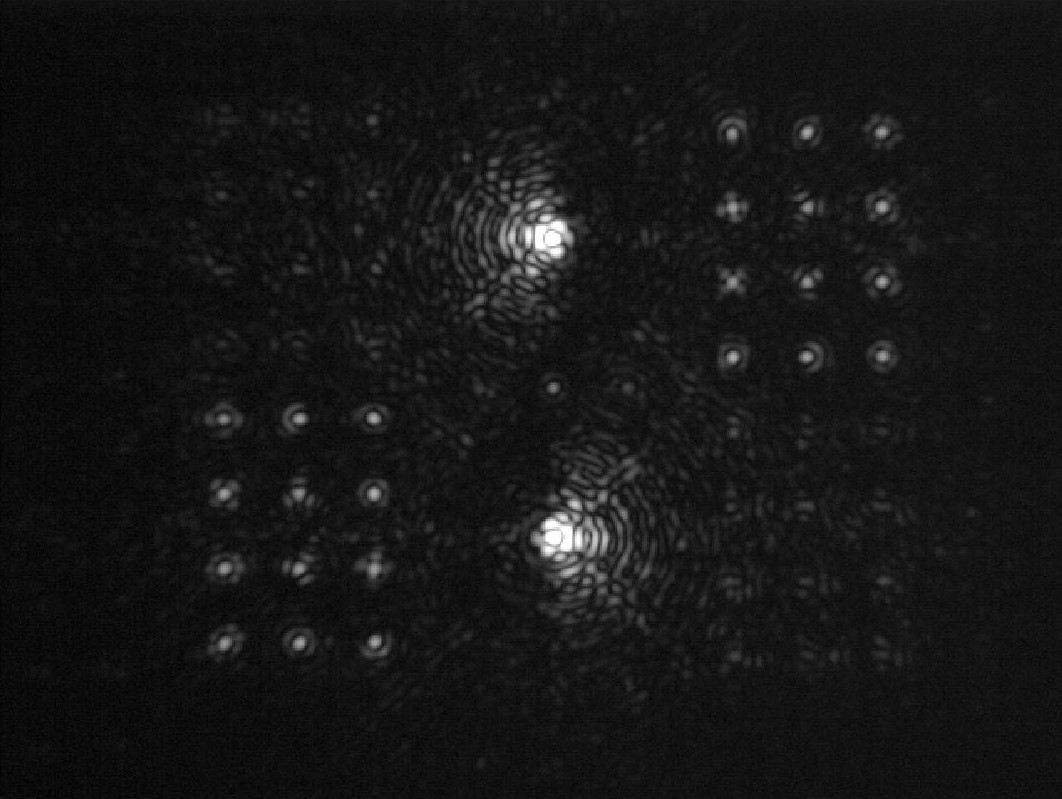}}
\caption{Comparison of the expected simulated vAPP (log scale) with the 12 Zernike MWFS and the image taken with the same mask in the laboratory (overexposed).}
\label{fig: lab_sim_comparison}
\end{figure}

\begin{figure}[H]
\centering
\includegraphics[scale=0.25]{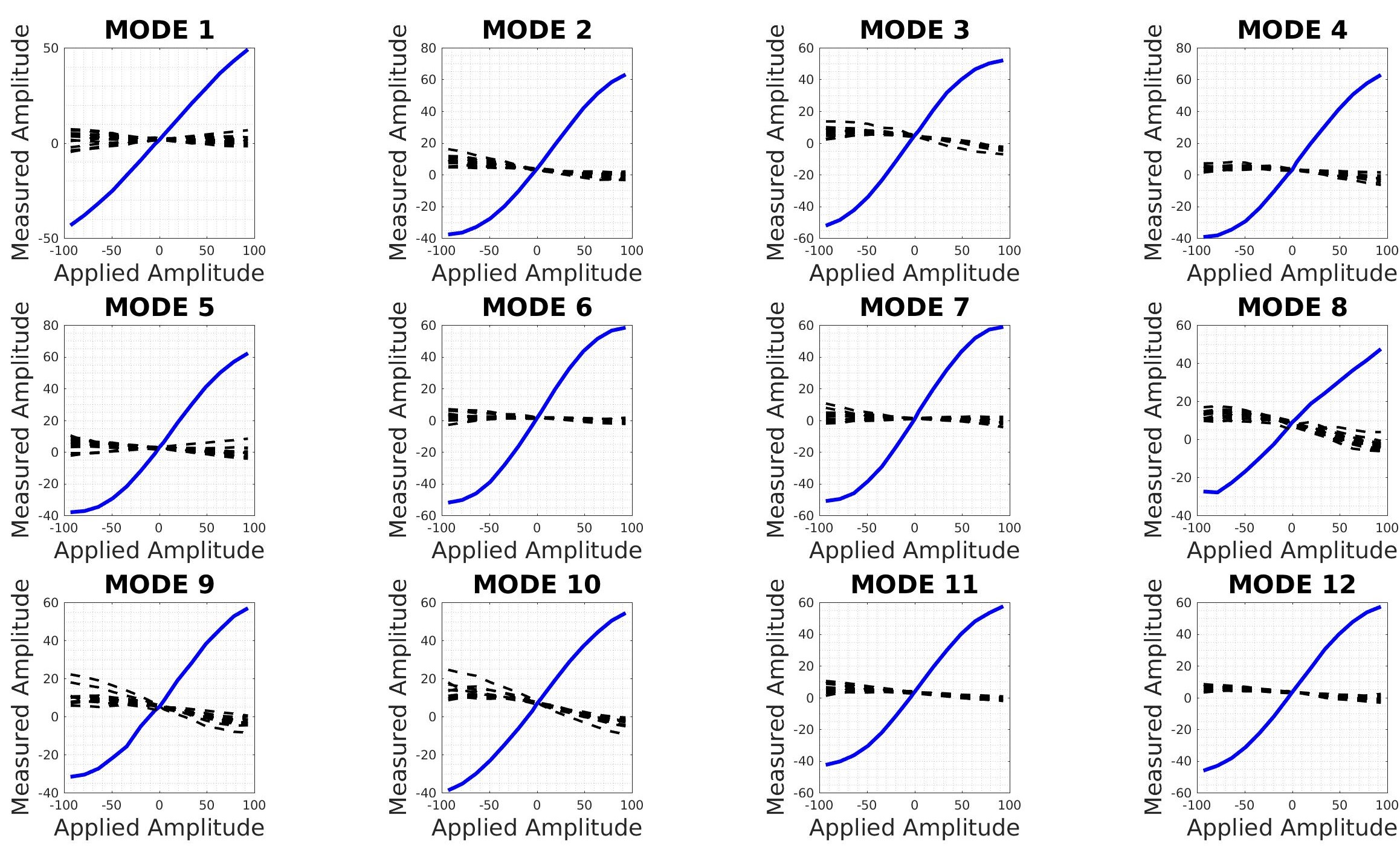}
\caption{Laboratory response curves showing the linear response of the MWFS to 12 Zernikes between +/- 100 nm amplitude aberrations.  The blue line represents the response of the mode applied, and the dashed black lines represent the response (crosstalk) of the 11 other modes to the applied mode. }
\label{fig: linearity_lab}
\end{figure}

For a direct comparison with the linearity tests performed in simulation, the twelve Zernike MWFS in the lab was tested by building a response matrix using the same wavefront sensor area cropping as in simulation and an aberration amplitude of 100 nm.  The same aberration amplitudes used in simulation were then applied, and the linear response of each mode and the subsequent crosstalk between modes were recorded in the plots shown in fig. \ref{fig: linearity_lab}.     \\

\begin{figure}[H]
\centering
\qquad
\subfloat[Response to testbed misalignment]{\includegraphics[scale=0.3]{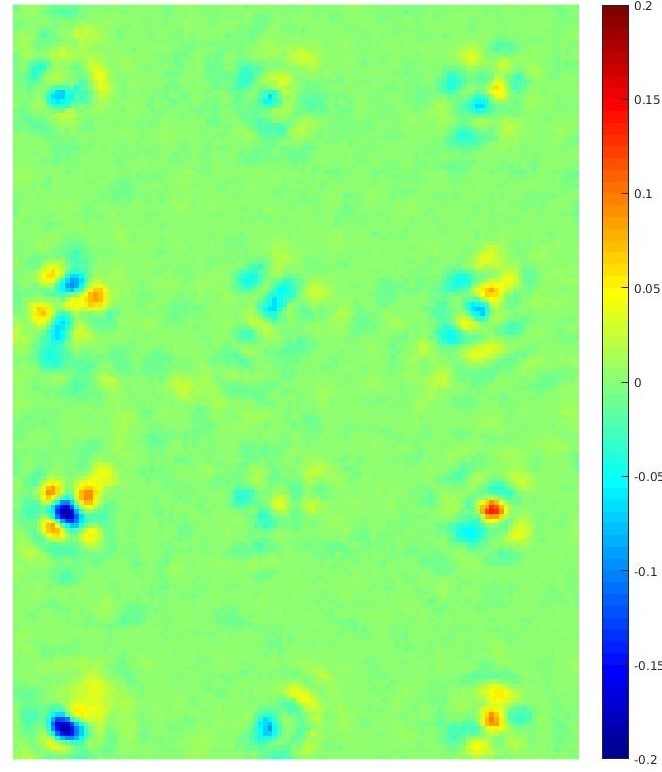}}
\qquad
\subfloat[Post-alignment correction]{\includegraphics[scale=0.3]{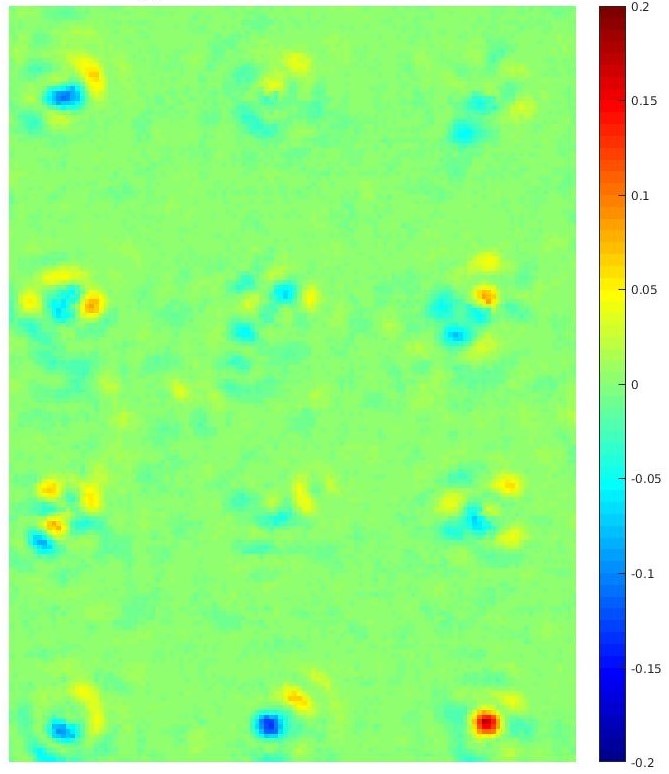}}
\caption{($\it{Left}$)  Defocus and astigmatism present on the UA Extreme Wavefront Control Lab testbed due to slight misalignment as seen by the vAPP 12 Zernike MWFS.  The two -0.2 amplitude peaks seen in blue corresponds to approximately -0.14 waves of defocus and oblique astigmatism.  ($\it{Right}$) After correction by the DM, the 12 Zernike MWFS shows the removal of the defocus and oblique astigmatism from the optical path but a slight increase in both vertical and horizontal coma.}
\label{fig: LOWFS_lab_aberration}
\end{figure}

Once again, the response of each mode to itself is plotted in blue, and the crosstalk between the other modes is shown by the dashed black lines.  It can be seen that the linear response regime determined in the lab for each mode is between +/-100 nm, thereby matching the results found in simulation.  However, the crosstalk between modes as seen in the lab demonstration has increased from the simulation results.  It is suspected that this is due both to noise and to slight misalignment in the optical path which induces astigmatism and/or coma upon reflection off the off-axis parabolic mirrors (OAPs).  The aberrations resulting from misalignment are sensed by the MWFS and are shown in fig. \ref{fig: LOWFS_lab_aberration}, where the MWFS response reveals the presence of defocus and oblique astigmatism. \\

\begin{figure}[H]
\centering
\quad
\subfloat[Uncorrected PSFs with astigmatism present]{\includegraphics[scale=0.25]{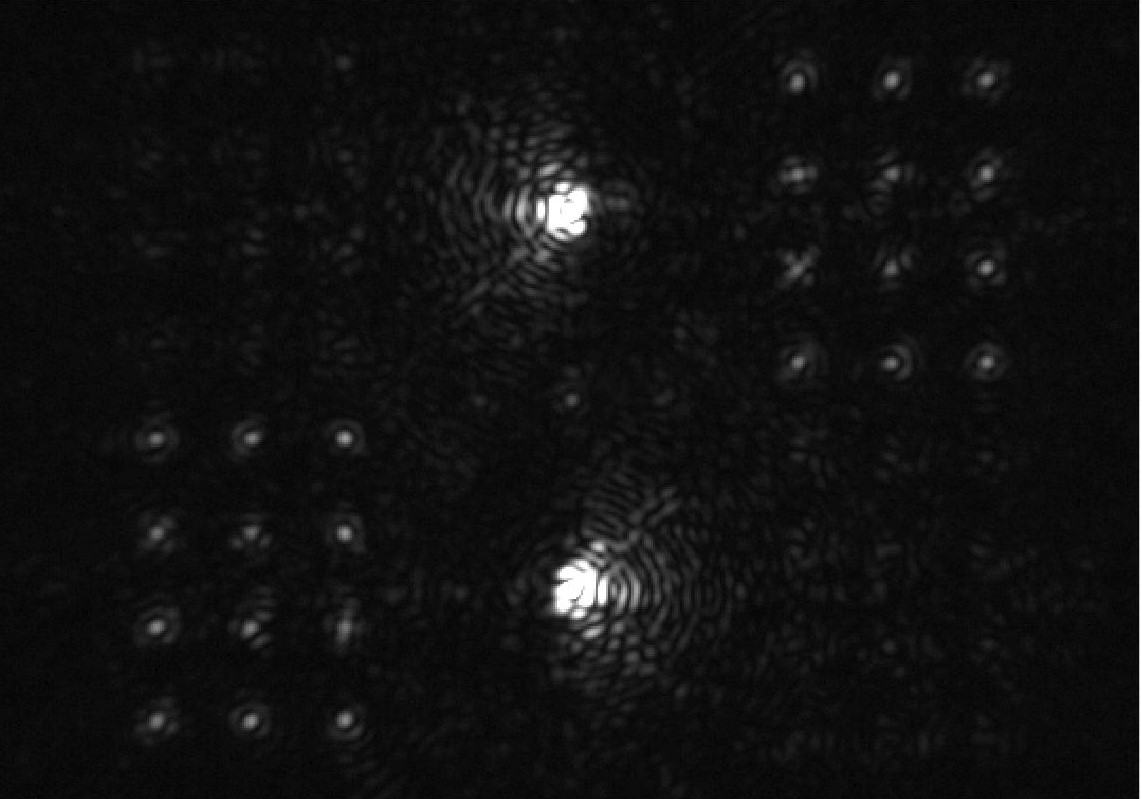}}
\quad
\subfloat[Corrected PSFs with astigmatism removed]{\includegraphics[scale=0.25]{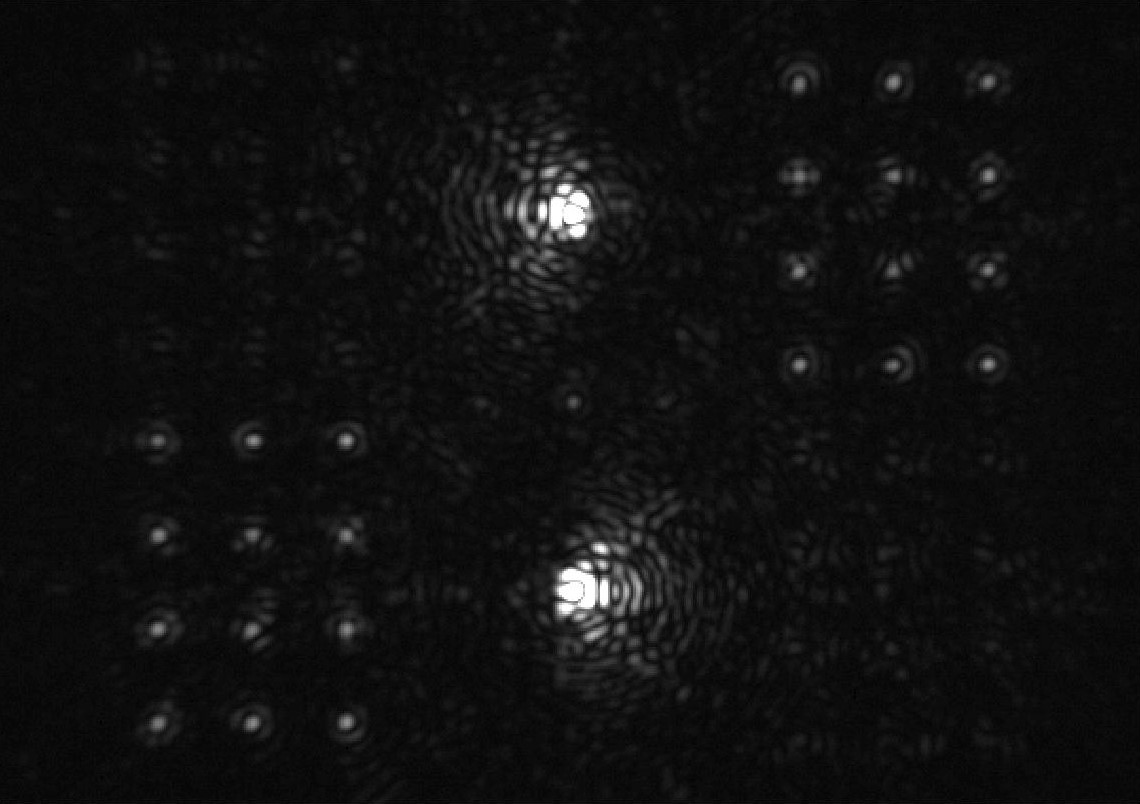}}
\caption{Lab results showing the DM removal of the oblique astigmatism and defocus sensed by the MWFS.}
\label{fig: alignment_correction}
\end{figure}

Using the 12 Zernike MWFS on the testbed as seen in fig. \ref{fig: LOWFS_lab_aberration}, the presence of approximately -0.14 waves of defocus and oblique astigmatism was detected in the optical path due to misalignment.  By applying +0.14 waves of both aberrations on the DM, the system alignment was corrected.  However, in doing so, some vertical and horizontal coma were also induced.  This effect is most likely due to the shifted position of the beam on the final OAP that occurs when applying astigmatism and defocus to the DM.  This misalignment on the OAP is the most likely source of the coma seen in the corrected image in fig. \ref{fig: LOWFS_lab_aberration}.  This interplay between astigmatism and coma is also suspected to be responsible for the greater crosstalk between modes seen in the lab results as compared to the expected modal crosstalk seen in simulation.  In spite of the induced coma, the astigmatism and defocus alignment correction increased the Strehl of the science PSFs which can be seen in fig. \ref{fig: alignment_correction} as an increase PSF core definition in both the upper and lower PSFs.  \\

\begin{figure}[H]
\centering
\includegraphics[scale=0.25]{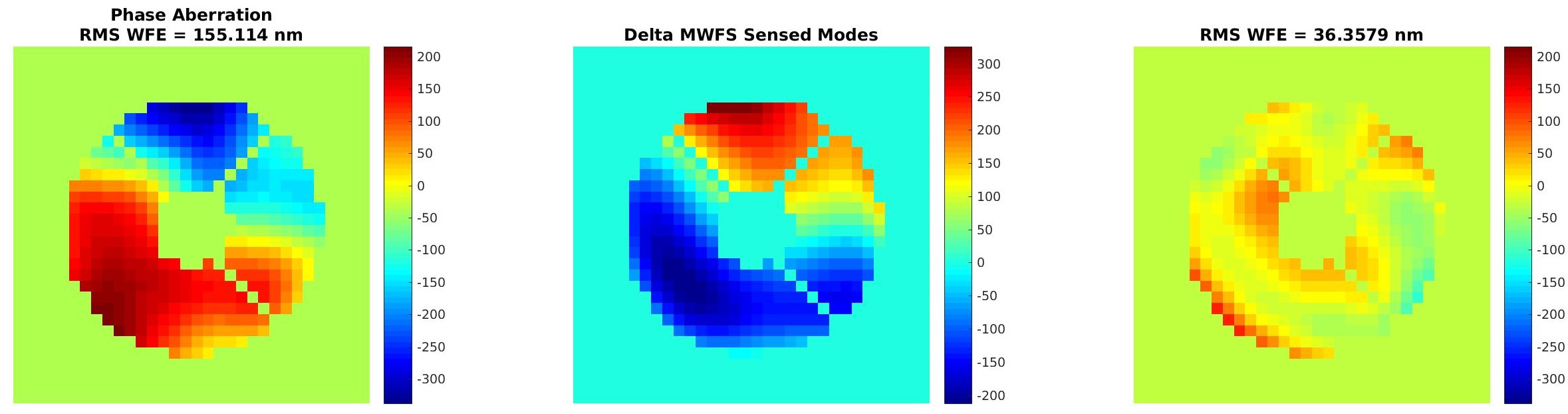}
\caption{LOWFS lab results using the 12 Zernike MWFS to sense an aberration applied on the BMC Kilo-DM and then corrected using the same DM.  ($\it{Left}$) The actuator displacement map of the low-order aberration applied to the DM with an RMS of 155 nm.  ($\it{Center}$)  The LOWFS-derived correction after 5 iterations applied to the DM.  ($\it{Right}$)  The residual wavefront error after correction by the DM with a final RMS of 36 nm.}
\label{fig: LOWFS_lab_pupil}
\end{figure}

\begin{figure}[H]
\centering
\includegraphics[scale=0.25]{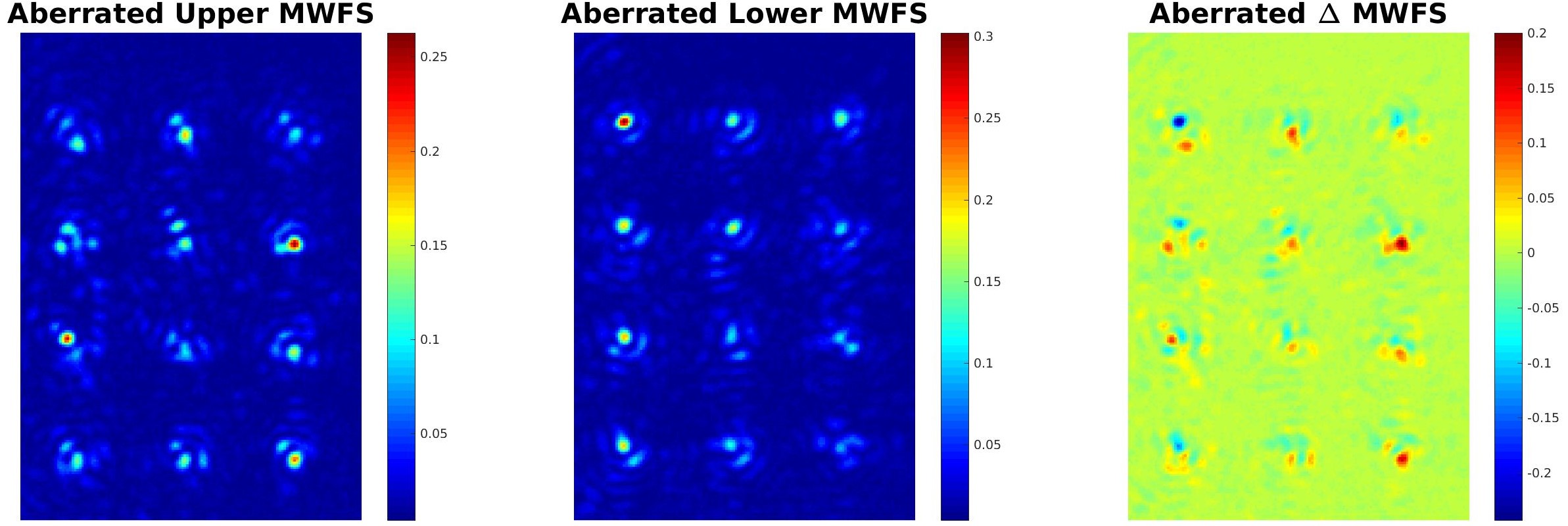}
\caption{The 12 Zernike MWFS response to the 155 nm aberration shown in fig. \ref{fig: LOWFS_lab_pupil}.   Upper MWFS PSFs ($\it{left}$), lower MWFS PSFs ($\it{center}$) and the difference of the two sets co-aligned ($\it{right}$) to produce the signal used for closed-loop LOWFS.}
\label{fig: LOWFS_lab_MWFS}
\end{figure}

\begin{figure}[H]
\centering
\quad
\subfloat[Log$_{10}$ upper vAPP science PSF]{\includegraphics[scale=0.4]{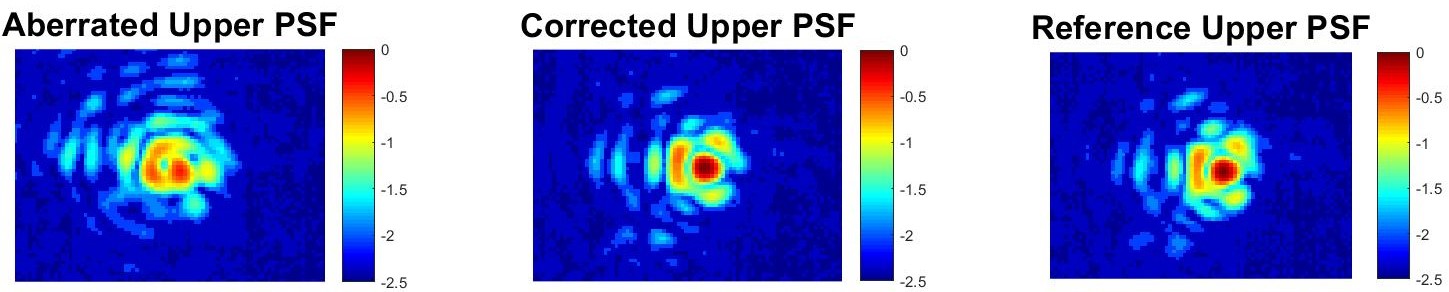}}
\quad
\subfloat[Log$_{10}$ lower vAPP science PSF]{\includegraphics[scale=0.4]{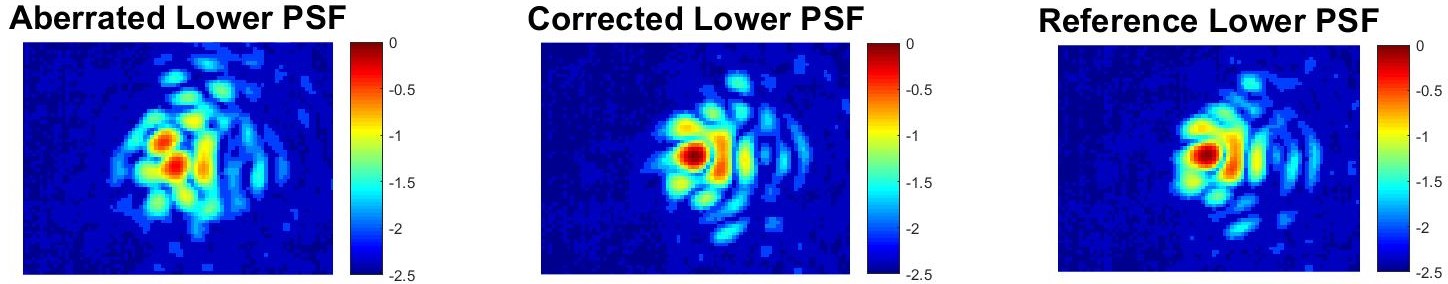}}
\caption{Lab results showing the science PSFs aberrated by the 155 nm RMS aberration shown in fig. \ref{fig: LOWFS_lab_pupil} and the LOWFS-driven correction derived by the 12 Zernike vAPP MWFS.  The unaberrated science PSFs are shown for reference.}
\label{fig: LOWFS_lab_PSFs}
\end{figure}

Following alignment correction, the MWFS was tested in closed-loop in the lab as it was in simulation.  The DM was used to inject an aberration into the beam path, and the derived correction was applied using the same DM.  Despite the increased crosstalk, the LOWFS loop converged, and in the case demonstrated in fig(s). \ref{fig: LOWFS_lab_pupil} - \ref{fig: LOWFS_lab_PSFs}, the initial aberration with an RMS of 155 nm was reduced to a residual error with an RMS of 36 nm after 5 iterations. \\

\section{Linear dark field control}\label{sec:LDFC}
\begin{figure}[H]
\centering
\includegraphics[scale=0.35]{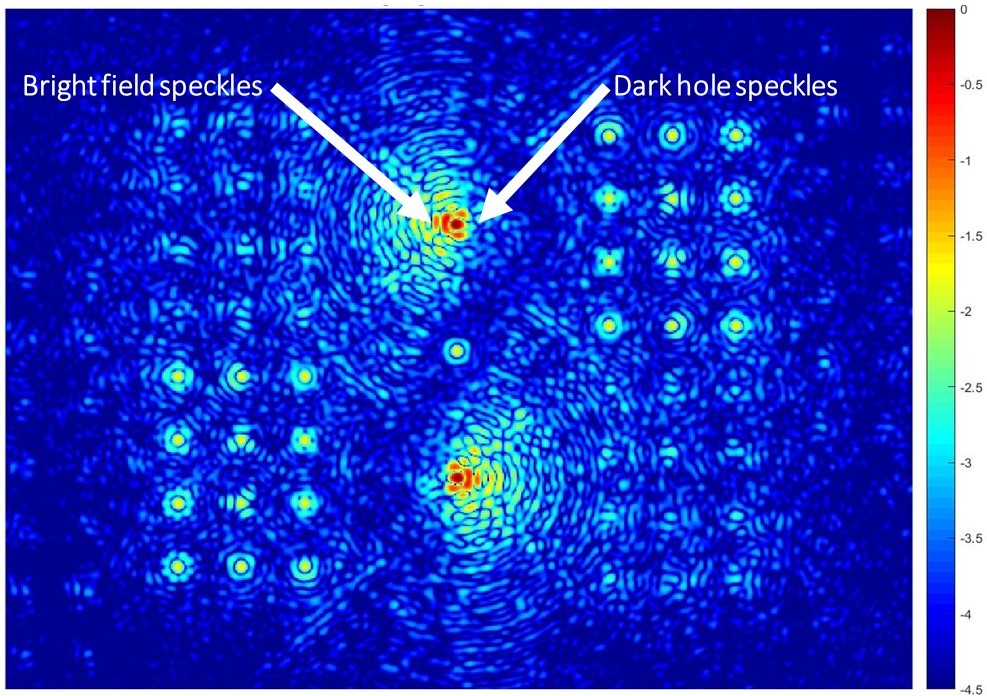}
\caption{The bright field speckles and the corresponding dark hole speckles induced by a mid-spatial frequency pupil plane aberration.  The bright field speckles are used to sense the aberration that is simultaneously corrupting the dark hole.}
\label{fig: LDFC_ref}
\end{figure}

While LOWFS/C controls low-order aberrations and maintains high Strehl, LOWFS/C is not capable of maintaining the high contrast in the dark hole.  The primary source of error that degrades the contrast within the dark hole is mid-spatial frequency aberrations.  To control these aberrations and to maintain the initial deep contrast delivered by the coronagraph, spatial linear dark field control (LDFC)\cite{Miller2017_JATIS_LDFC} is employed.  LDFC is a similar algorithm to LOWFS in that it provides a relative wavefront error measurement rather than an absolute phase measurement like electric field conjugation or similar techniques.\cite{Groff2015_FPWFS}  To sense the aberrations that degrade the dark hole, spatial LDFC measures the relative changes in intensity of the bright field within the same spatial frequency extent as the dark hole but on the opposite side of the stellar PSF as seen in fig. \ref{fig: LDFC_ref}  \\

\begin{figure}[H]
\centering
\includegraphics[scale=0.35]{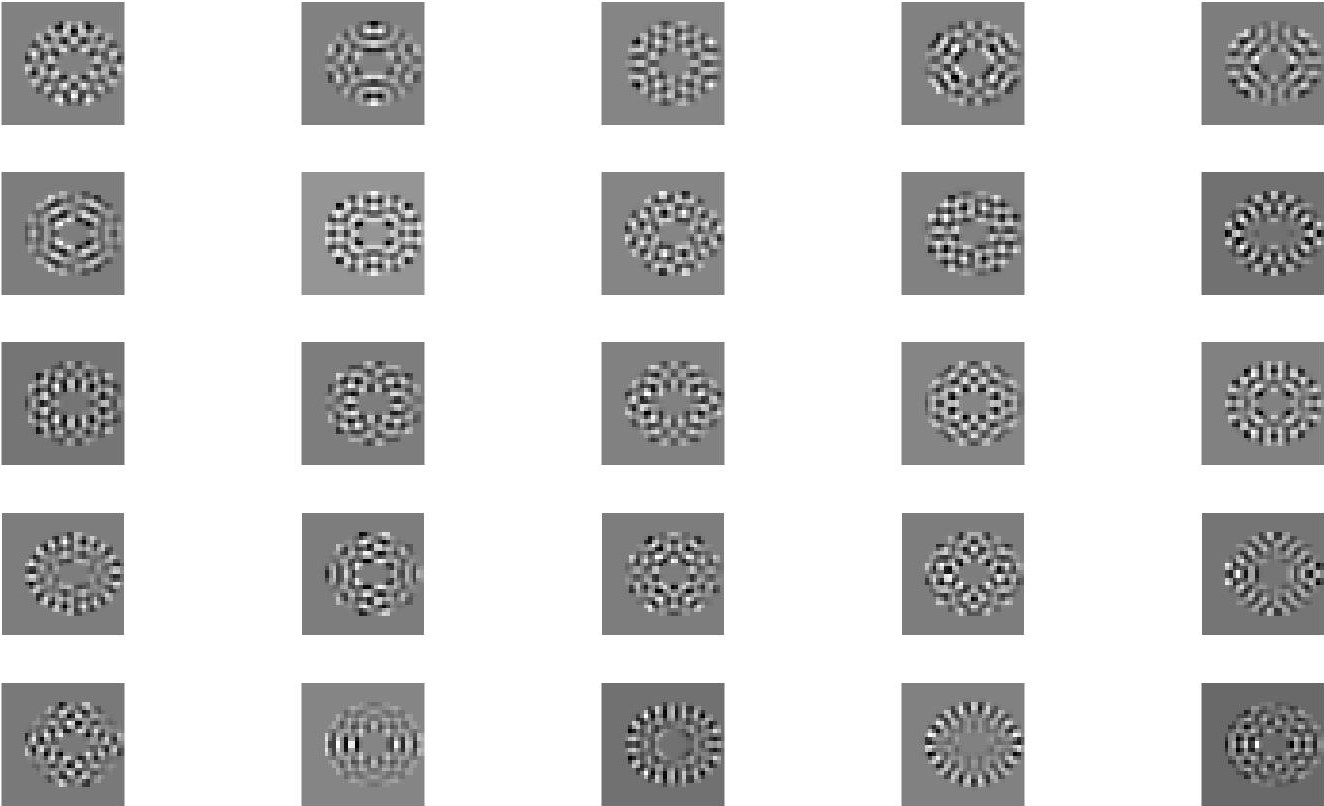}
\caption{25 of the 100 mid-spatial frequency modes used in the LDFC control loop }
\label{fig: LDFC_modes}
\end{figure}

Bright field speckles of a high enough magnitude respond linearly to aberrations in the pupil plane.\cite{Miller2017_JATIS_LDFC}  This monotonic response allows for closed-loop control of both the bright field and dark hole speckles induced by the same pupil plane aberration.  To build this closed loop, rather than using low-order modes, mid-spatial frequency modes derived from the influence functions of the DM are used as the modal basis set.  These modes are chosen such that the spatial frequency content of the basis set matches the spatial extent of the dark hole.  For the following tests in both simulation and in the lab, 100 modes are used to control the dark hole.  The first 25 of these modes are shown in fig. \ref{fig: LDFC_modes}.     \\

In the following work in both simulation and in the lab, the spatial extent of the dark hole is 2 - 15 $\lambda$/D.  The control radius however, is set by the number of illuminated actuators across the DM which, in this case, is 22, thereby setting the greatest controllable spatial frequency to 11 $\lambda$/D (fig. \ref{fig: LDFC_in_simulation}).   \\

\subsection{LDFC in simulation}
\begin{figure}[H]
\centering
\qquad
\subfloat[Defocused vAPP image]{\includegraphics[scale=0.25]{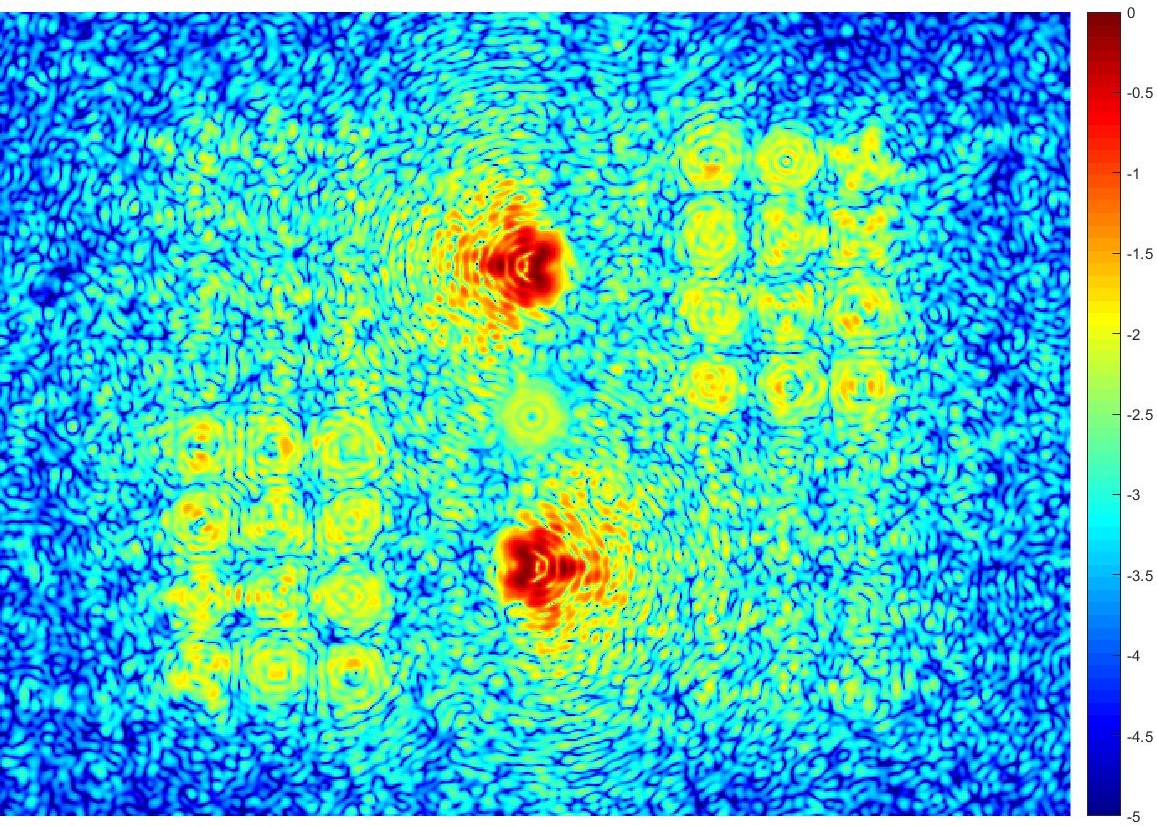}}
\qquad
\subfloat[Bright pixels selected for the LDFC response matrix]{\includegraphics[scale=0.1]{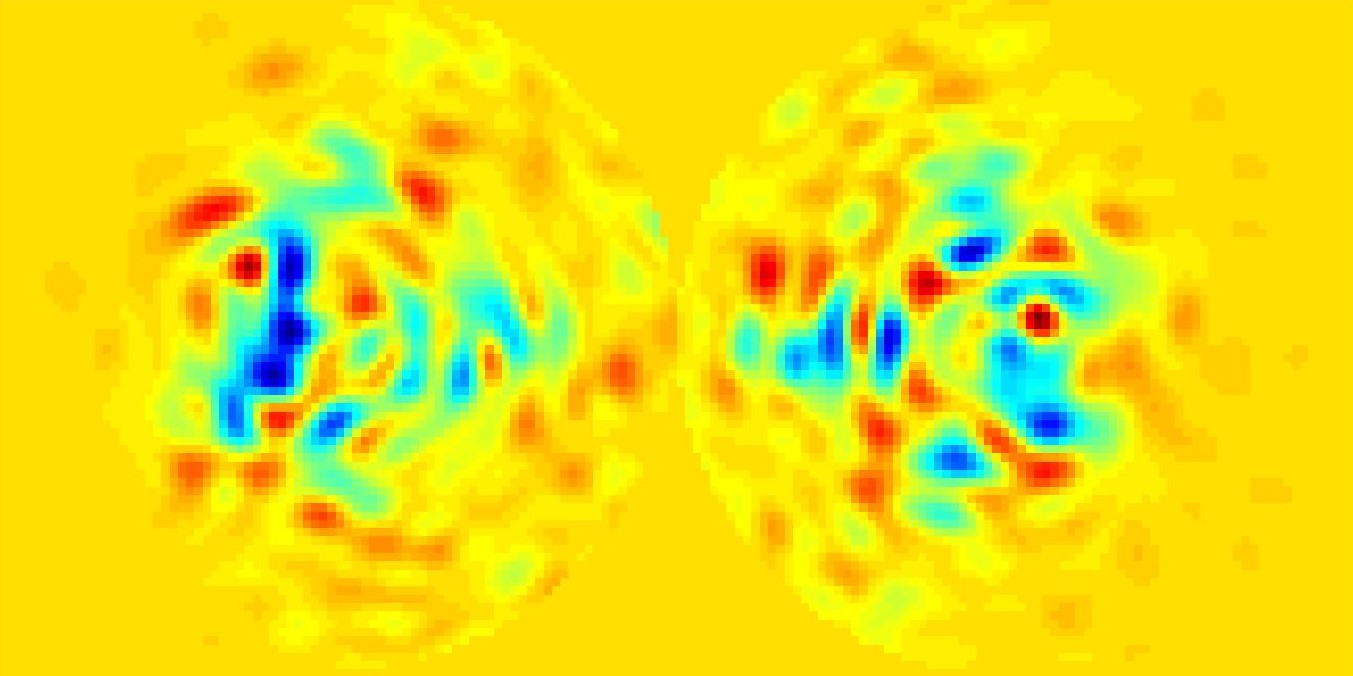}}
\caption{The defocused vAPP image used to provide sign information for the LDFC control loop.}
\label{fig: LDFC_defocused}
\end{figure}

In simulation, to provide sign information, the image is defocused as shown in fig. \ref{fig: LDFC_defocused}.  The bright pixels within the control radius of the DM are then selected and used in the response matrix for the control loop.  The response matrix is built by recording the intensity change of these bright pixels to the 100 mid-spatial frequency modes chosen to match the controllable extent of the dark hole.    \\

\begin{figure}[H]
\centering
\includegraphics[scale=0.4]{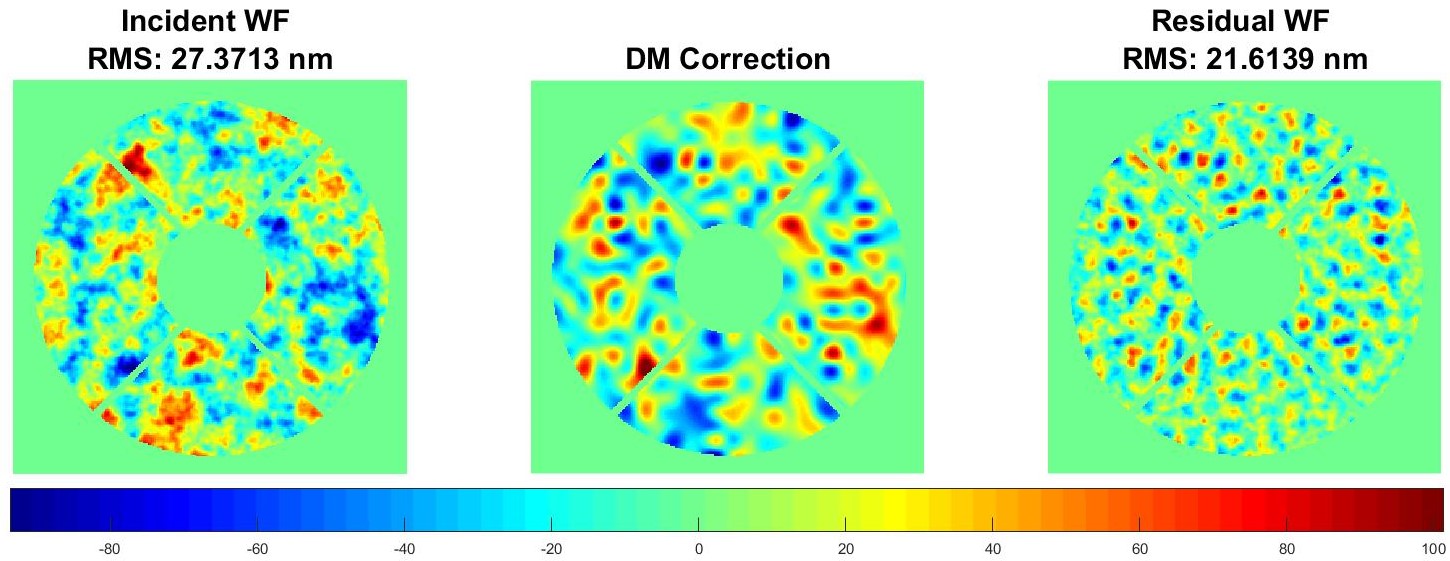}
\caption{LDFC in simulation reducing the residual RMS wavefront error from 27 nm to 21 nm.}
\label{fig: LDFC_demo}
\end{figure}

\begin{figure}[H]
\centering
\includegraphics[scale=0.45]{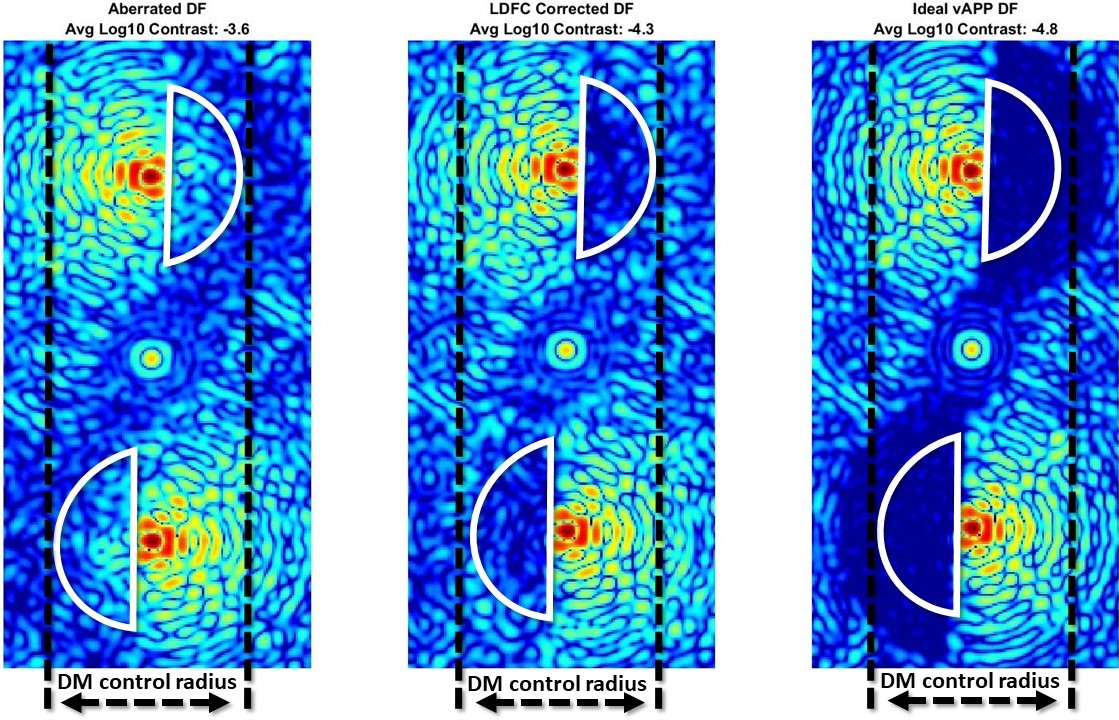}
\caption{Dark hole correction using linear dark field control in simulation.  While the OWA of the dark hole is 15 $\lambda$/D, the control radius set by the DM is 11 $\lambda$/D.}
\label{fig: LDFC_in_simulation}
\end{figure}

To demonstrate LDFC in simulation, a pupil plane aberration containing mid-spatial frequencies between 2 - 15 $\lambda$/D was injected.  The RMS of the injected aberration in the following example was 27 nm (fig. \ref{fig: LDFC_demo}).  After running LDFC, the RMS of the residual wavefront error was reduced to 21 nm.  This reduction in wavefront error reduced the average dark hole contrast across the entire 2 - 15 $\lambda$/D extent from $10^{-3.6}$ back down to $10^{-4.3}$ (fig. \ref{fig: LDFC_in_simulation}).    

\subsection{LDFC laboratory demonstration}
\begin{figure}[H]
\centering
\includegraphics[scale=0.30]{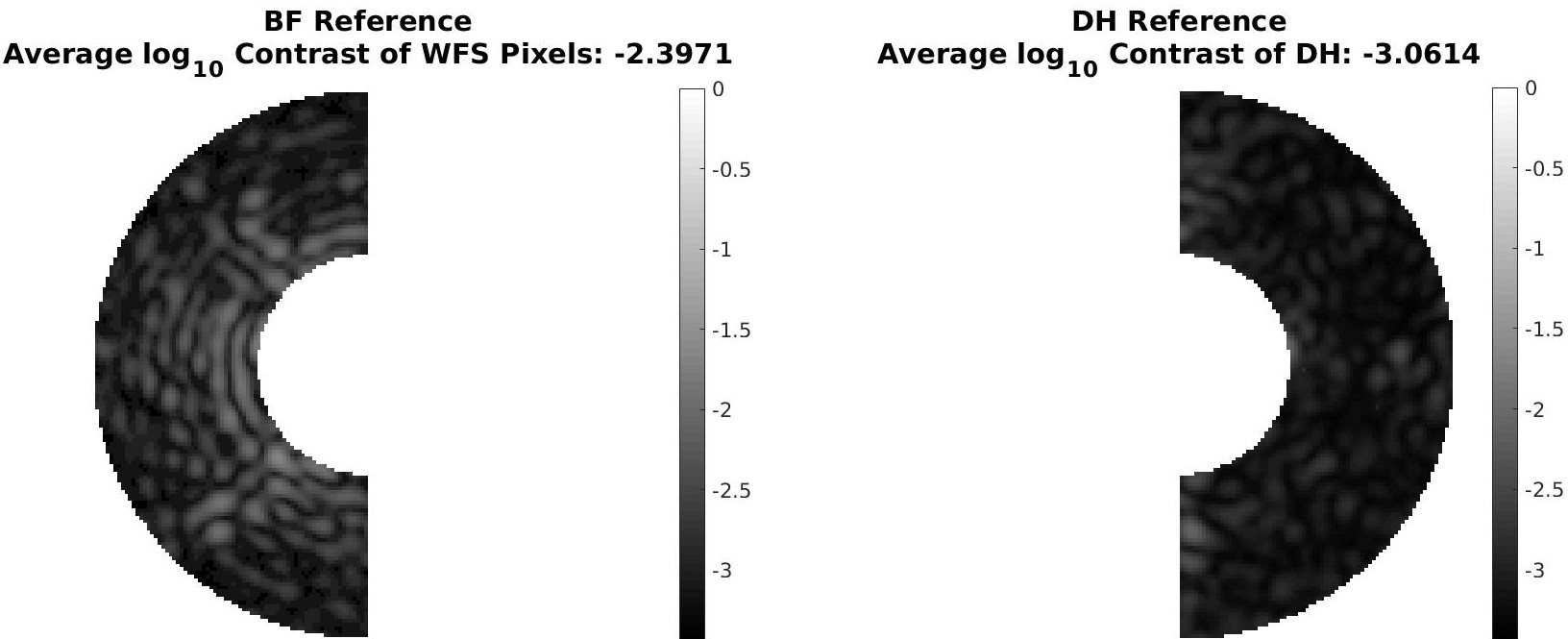}
\caption{Bright field and dark hole lab unaberrated reference images}
\label{fig: LDFC_lab_references}
\end{figure}

In the lab, aberrations due to optical surface errors limit the initial vAPP contrast to approximately $10^{-3}$.  The average contrast of the bright field used in the LDFC response matrix was $10^{-2.4}$ (fig. \ref{fig: LDFC_lab_references}).  In the following preliminary lab results, the signal used for LDFC was not defocused.  This limited the algorithm's ability in some cases to determine the correct sign of the aberration; this sign ambiguity can be dealt with by defocusing the signal as was done in simulation.   \\

\begin{figure}[H]
\centering
\includegraphics[scale=0.35]{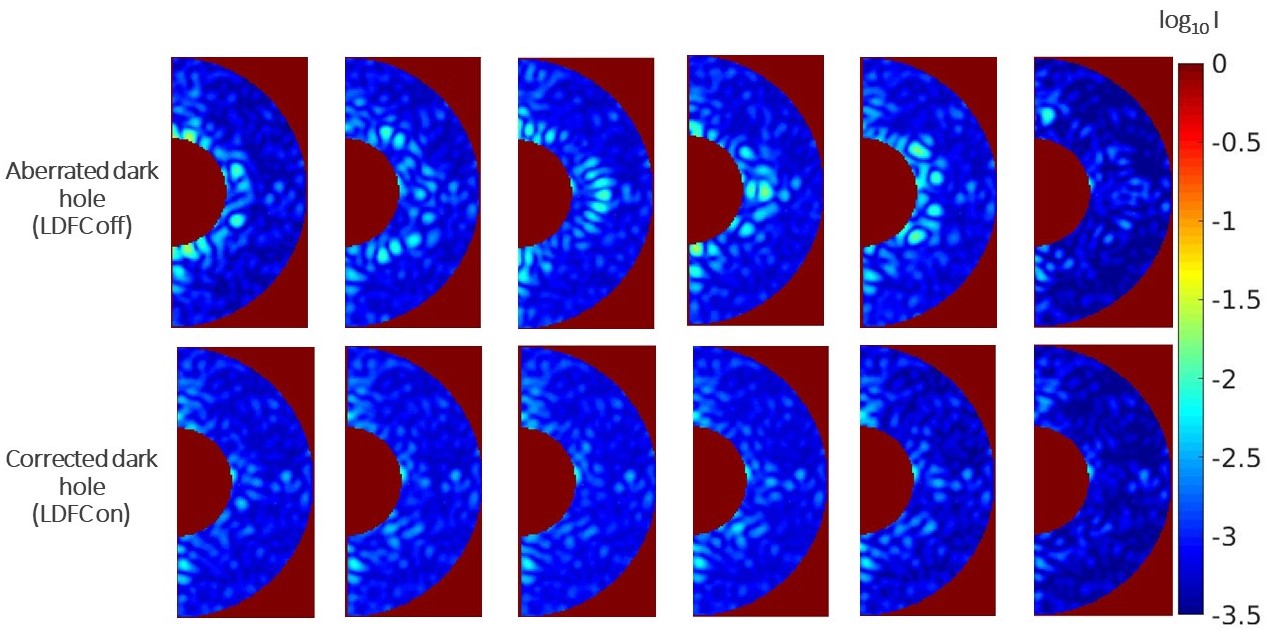}
\caption{Dark hole correction by LDFC in the lab, decreasing $10^{-2}$ magnitude speckles back to the intial $10^{-3}$ contrast floor}
\label{fig: LDFC_in_the_lab}
\end{figure}

The following results were obtained by applying aberrations on the DM that were a linear combination a random number of the modes used in the LDFC response matrix.  The magnitude of the speckles produced by these aberrations was on the order of $10^{-2}$.  LDFC was then used to sense those aberrations in the bright field, and the appropriate correction was applied on the DM to return the dark hole to its initial dark hole contrast of $10^{-3}$ (fig. \ref{fig: LDFC_in_the_lab}).   

Further testing of LDFC in the lab will include injection of random phase aberrations and defocusing the image to retrieve sign information.  \\
\section{Future on MagAO-X and Conclusions}
\begin{figure}[H]
\centering
\includegraphics[scale=0.35]{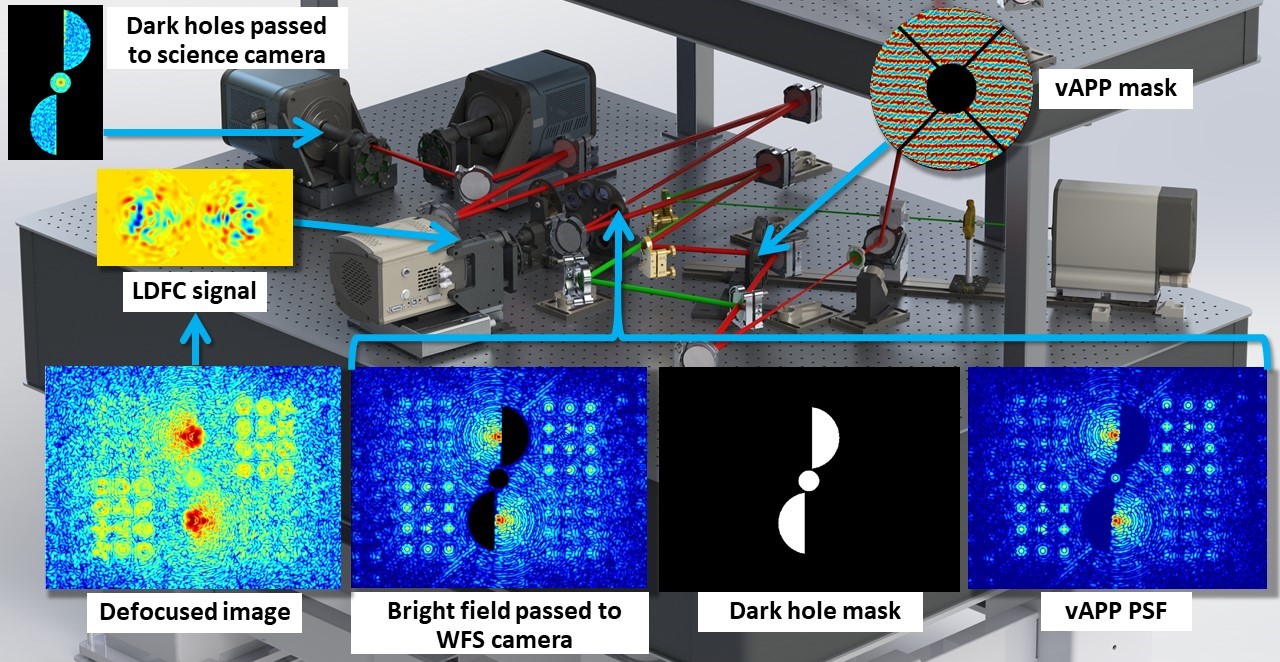}
\caption{Future layout for LDFC and LOWFS using MWFS PSFs on MagAO-X \cite{Close2017_MagAOX}}
\label{fig: LDFC_on_MagAOX}
\end{figure}

On MagAO-X, the light used by both LOWFS and LDFC will be separated from the science channel by a binary mask, placed at an intermediate focal plane, which transmits the dark holes to the science camera and reflects the stellar bright field back to a dedicated WFS camera shown in fig. \ref{fig: LDFC_on_MagAOX}.  This reflected light will contain both the MWFS PSFs and the bright field used by LDFC.  This bright field signal can then be used to run LOWFS and LDFC simultaneously using different regions of the same image as the wavefront sensor for both algorithms.  Running both algorithms in the science image will allow for the maintenance of high Strehl as well as high dark hole contrast.     \\

Testing of the vAPP MWFS and LDFC is ongoing at the UA Extreme Wavefront Control Lab.  The official design of the MagAO-X design is currently being finalized, and the null space of LDFC is under investigation.  These efforts will inform the expected performance of both LOWFS with the vAPP MWFS and LDFC on the MagAO-X instrument which will see first light in March of 2019.
\pagebreak
\acknowledgments     
This work was supported [in part] by NSF MRI Award  $\#$1625441 (MagAO-X).

\bibliography{DissertationBib}
\bibliographystyle{spiebib}

\end{document}